# GaSbBi alloys and heterostructures: fabrication and properties

O. Delorme, L. Cerutti, R. Kudrawiec, E. Luna, J. Kopaczek, M. Gladysiewicz, A. Trampert, E. Tournié, and J.-B. Rodriguez

**Abstract:** Dilute bismuth (Bi) III-V alloys have recently attracted great attention, due to their properties of band-gap reduction and spin-orbit splitting. The incorporation of Bi into antimonide based III-V semiconductors is very attractive for the development of new optoelectronic devices working in the mid-infrared range (2 – 5 µm). However, due to its large size, Bi does not readily incorporate into III-V alloys and the epitaxy of III-V dilute bismides is thus very challenging. This book chapter presents the most recent developments in the epitaxy and characterization of GaSbBi alloys and heterostructures.

O. Delorme, L. Cerutti, E. Tournié, J.-B. Rodriguez*
IES, Univ. Montpellier, CNRS,
34000 Montpellier, France

R. Kudrawiec, J. Kopaczek, M. Gladysiewicz
Faculty of Fundamental Problems of Technology, Wrocław University of Science and Technology,
Wybrzeże Wyspiańskiego 27, 50-370 Wrocław, Poland

E. Luna, A. Trampert
Paul-Drude-Institut für Festkörperelektronik, Leibniz-Institut im Forschungsverbund Berlin e.V.,
Hausvogteiplatz 5-7, D-10117, Berlin, Germany

*Corresponding author: jean-baptiste.rodriguez@umontpellier.fr

# 1. Introduction: motivation and historical overview

## 1.1. The antimonides

The antimonides are the material system comprising GaSb, InAs, AlSb III-V semiconductors, as well as their associated ternary, quaternary or quinary alloys, *e.g.* AlGa(In)AsSb or GaInAsSb. They are usually fabricated using Molecular Beam Epitaxy (MBE) on GaSb or InAs high quality substrates. The particularity of these materials lies in the large variety of alloys achievable, spanning a very wide range of band-gaps, allowing the realization of quantum-well (QW) lasers operating from the near [Cerutti-2015] to the mid-infrared [Hosoda-2010] (Fig. 1). This material system is also unique in terms of band-offsets: InAs has a type-III band alignment with GaSb for example, with the maximum of the valence band of GaSb lying at a higher energy than the minimum of the conduction band of InAs (Fig. 1). This semi-metallic interface is particularly interesting, since it allows fabricating superlattices having a fundamental transition energy as small as desired [Wei-2002, Rhiger-2011, Tan-2018] or very efficient carrier transfer from the valence band to the conduction band. This original feature has been successfully used to design high-performance photodetector arrays covering the whole infrared range [Razeghi-2014], but is also at play in topological insulator structures and Inter-band Cascade Lasers (ICLs) for example [Liu-2008, Vurgaftman-2015].

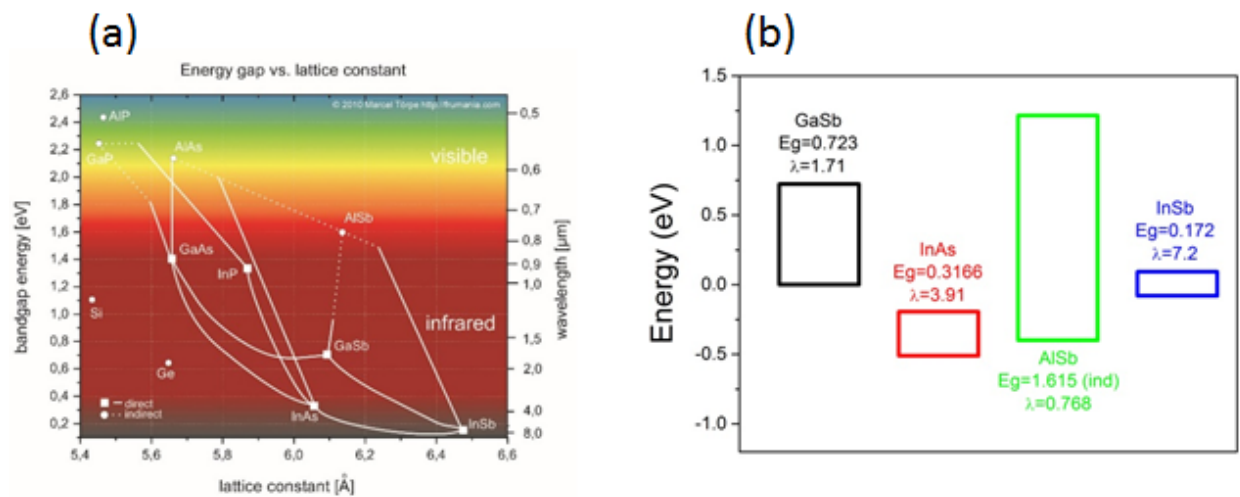

**Fig. 1** (a) Bandgap versus lattice-constant for various semiconductor materials. Reprinted with permission from http://lab.frumania.com/wp-content/uploads/2010/06/bandgap_edit.jpg (b) The band alignment between the different antimonide materials, GaSb and InAs have a type-III alignment.

Additionally, InAs and AlSb have one of the largest conduction band offset achievable with semiconductor materials, which stemmed the development of quantum-cascade lasers (QCLs) from the mid-wavelength infrared range [Lafaille-2012] to the far infrared range [Bahriz-2015] using these two binaries.

Beyond the natural use of antimonides for infrared optoelectronic, a lot of research has also been conducted on high-speed/low-consumption electronics with these materials, because of the very large mobilities and narrow band-gaps achievable [Gardes-2014]. Finally, it was recently shown that the antimonides are serious candidates for the direct integration of III-V semiconductor on silicon [Reboul-2011, Castellano-2017, Nguyen-Van-2018].

All these assets are at the origin of a still intense research held by a worldwide community. The applications covered by devices made from antimonides span a wide range of activities, from trace-gas detection and gas spectroscopy [Willer-2006, Tittel-2013] to night-vision [O'Malley-2010], astronomy [Winnewisser-1994, Abedin-2006], non-invasive medical diagnosis [Godoy-2015, Gurjarpadhye-2015] and strategic military applications [Tidrow-2001].

## 1.2. Motivation for the development of GaSbBi

Type-I quantum-well (QW) lasers in the mid-IR have for a long time been one of the main driving force for the development of the Sb-based technology [Tournié & Baranov, 2012]. Such lasers, with GaInAsSb QWs embedded in AlGaAsSb barriers allowed the realization of room-temperature (RT), continuous-wave (CW) operating devices exhibiting low-threshold current and high output power between 1.9 to 2.9 µm [Chen-2008]. More recently, the wavelength range has been extended to the near-IR [Cerutti-2015], down to telecom wavelengths (1.5 µm), which paves the way for new applications for this material system. To this end, composite QWs, based on a stacking of GaInSb and AlInSb layers was employed in order to independently control the strain and the thickness of the QW. Increasing the emission wavelength above 3 µm is also a very hot topic, since many important gases have absorption lines there [Rothman-2009]. By incorporating Indium in the barrier layers, some labs managed to demonstrate RT/CW lasers above 3 µm [Hosoda-2010]. Indeed, AlGaAsSb/GaInAsSb QWs does not allow efficient emission above 2.9 µm because the hole confinement decreases rapidly with the wavelength (Fig. 2). Adding In into the barriers improves to some extent this confinement and laser emission could be obtained up to around 3.5 µm. The laser performance however severely degrades when the wavelength increases beyond 3 µm [Hosoda-2010]. Additionally, the control over the realization of such a quinary material (AlGaInAsSb) is a complex task and new solutions are to be found in order to further increase the emission wavelength. Interband Cascade Lasers (ICLs) are a promising candidate for making lasers emitting above 3 µm, due in particular to lower Auger losses [Vurgaftman-2015]. However, the epitaxy complexity remains an even more severe challenge, and the radiative transition occurs in a type-II configuration which tends to lower the gain when increasing the emission wavelength. Quantum Cascade lasers (QCLs), relying on an intra-band radiative transition, are now well developed but the typical emission wavelength of these lasers is beyond 5 µm [Pecharroman-2013]. In summary, a simple and efficient way of making high performance lasers in the 3-5 µm range is still missing, despite the large number of applications (gas detection, countermeasure, etc…) that could benefit from such devices.

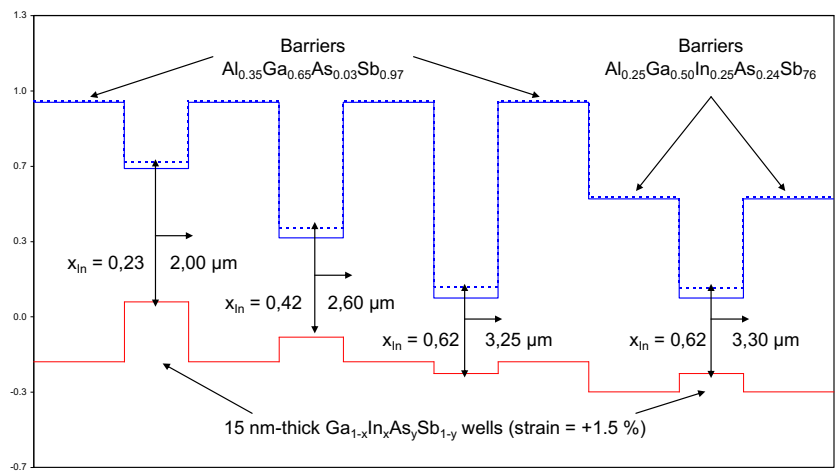

**Fig. 2** Conduction (blue) and valence (red) band alignment of antimonide type-I QWs emitting in the mid-IR. AlGaAsSb/GaInAsSb QWs allow very effective light emission in the two-first cases. For emission above 3 µm however (third configuration), the increase of the In content in the QW material leads to a configuration where the holes are not anymore confined in the QW. Replacing the AlGaAsSb barrier material by the quinary alloy (AlGaInAsSb) allows maintaining to some extend the hole confinement and thus increase the emission wavelength, but at a high cost in terms of complexity.

Incorporating Bi into antimonide alloys is extremely promising in terms of new device design possibilities. On the one hand, the lattice parameter of GaBi is close to 6.3 angstrom [Janotti-2002,

Tixier-2003, Rajpalke-2014] making it possible to grow coherently strained layers on GaSb. On the other hand, there is an impressive difference between the band-gaps of GaSb and GaBi (estimated between -2.91 eV and -1.45 eV) [Janotti-2002, Ferhat-2006, Samajdar-2014], and any antimonide alloy comprising Bi would thus see its electronic properties drastically modified. Among them, GaSbBi is an alloy that has a strong potential for the realization of type-I QW lasers emitting above 3 µm. Indeed, the heterostructure formed by GaSb and GaSbBi is Type-I (Fig. 3), allowing the realization of efficient quantum wells emitting in the whole mid-wave IR domain. Advantages of such lasers compared to the state-of-the-art devices (inter-band cascade lasers (ICLs) or lasers with type-I quinaries QW barriers) in the mid-wave IR are listed below and illustrated on Fig. 3:

- Simpler design,
- GaSb QW barriers allow a better heat dissipation than the quaternary or quinary alloys,
- The hole confinement improves as the Bi content (and therefore the emission wavelength) is increased,
- The spin-orbit splitting is drastically increased by the adjunction of Bi [Sweeney-13], leading to a decrease of the Auger losses,
- The active region is aluminum-free, which is known to be beneficial for the device lifetime

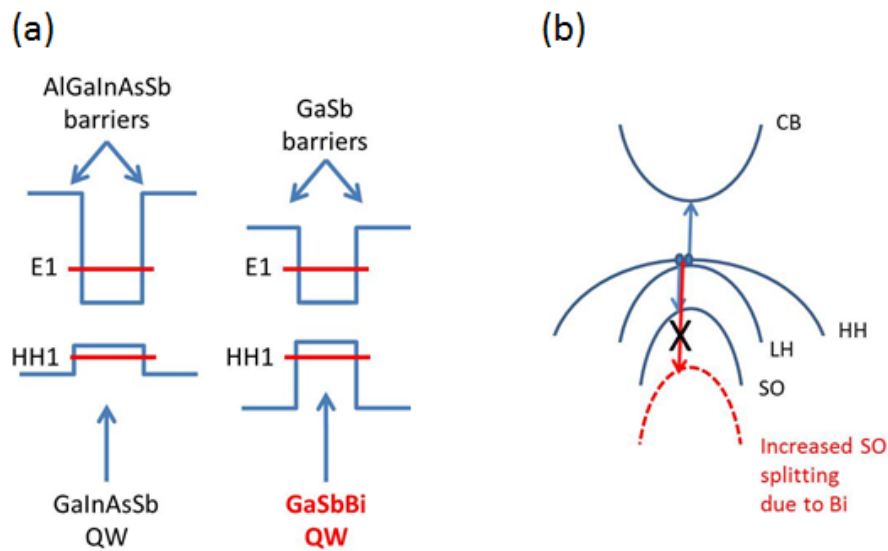

**Fig. 3** (a) standard type-I QW for emission beyond 3 µm are complex and so far limited to an emission wavelength ~3.5 µm, using GaSbBi as the QW material drastically reduce the complexity and allows the use of Al-free barriers. (b) The use of Bi in the QW allows reaching wavelength not achievable with common antimonides thanks to the reduction of Auger losses due to the resonance between the bandgap energy and the split-off band energy, by increasing significantly of the latter.

## 1.3. Historical overview

The idea of using Bi with the standard III-V semiconductors has been around since 1972, when Joukoff *et al.* [Joukoff-1972] reported the growth of bulk InSbBi crystals by the Czochralski method. Unfortunately the solubility of Bi on the substitutional sites in InSbBi is limited to 2.2% for equilibrium crystal growth techniques, because InBi has a tetragonal lattice when InSb has a cubic lattice. Later on, non-equilibrium growth techniques such as MBE [Noreika-1982] or MOCVD [Wagener-2000] enabled to obtain metastable films with increased Bi solubility. But the Bi composition obtained at the time remained nevertheless rather small ~3% and the sample surface was covered with Bi droplets.

Following these somewhat disappointing experiments, Bi was rather used as a surfactant to improve the quality of GaAs(N) dilute nitrides [Tixier-2003a]. Indeed, the Bismuth is a large atom which tends to not incorporate into the growing film under typical growth conditions, but rather segregates to the surface where its presence modifies the kinetics of the other impinging atoms. Such effect has been used to improve the luminescence efficiency and to smooth the surface of GaNAs layers for example [Tixier-2003a]. The successful use of Bi as a surfactant encouraged new experiments regarding its incorporation into GaAs. Thanks to drastic deviations from typical GaAs growth condition in MBE or MOCVD [Tixier-2003b, Oe-2002, Yoshimoto-2003], the maximum incorporation of Bi was increased up to 4.5% in GaAs, using both a low growth temperature (~360°C) and a low V/III flux ratio. More recently, a careful control of the Ga/As flux ratio combined with a growth temperature lower than 300°C allowed to incorporate 22% of active Bi in the layer [Lewis-2012]. If research on GaAsBi has mainly been devoted to the realization and properties of thick layers, some investigations on GaAsBi QWs can also be found in the literature. For example, the photoluminescence (PL) efficiency from GaAsBi QWs with 7% of Bi and emitting at 1.17 µm [Makhloufi-2014] was improved thanks to thermal annealing treatment, contributing to decrease the density of localized defects due to Bi aggregates or alloy disorders.

In 2013, in the framework of the European project BIANCHO [BIANCHO-2010], laser emission at RT under pulsed operation was demonstrated for the first time from a structure comprising GaAs$_{0.98}$Bi$_{0.02}$ QWs [Ludewig-2013]. Although the threshold density was high, the emission wavelength was longer than with GaAs QWs. In 2014, the University of Kyoto fabricated a laser structure by MBE with 4% Bi in the GaAs QWs [Fuyuki-2014]. Pulsed lasing emission at 1.045 µm (RT) was obtained, with reduced thermal tunability. In 2015, GaAsBi-based MQW lasers with Bi content up to 8% were grown by MBE and MOVPE, with a maximum lasing emission of 1.06 µm and a threshold current density of 12.5 kA/cm$^2$ at RT [Marko-2015]. Temperature- and pressure dependent measurements of stimulated and pure spontaneous emission measurements showed that this high threshold density was caused by non-radiative defect-related recombination and an inhomogeneous carrier distribution. This was attributed by the authors to the inhomogeneous QW width and non-uniform Bi composition. More recently, a GaAsBi/GaAs laser diode with 5.8% Bi was realized, with emission up to 1.142 µm at RT. For the first time, continuous wave lasing was demonstrated, up to 273 K [Wu-2017]. Even if most investigations currently focus on GaAsBi, the incorporation of Bi in other III-V semiconductors has also been attempted. In 2014, a layer of InPBi$_{0.024}$ grown by gas source MBE on an InP substrate presented a band-edge absorption at 1.05 µm, representing a redshift of 120 nm with respect to InP [Gu-2014]. In addition, a photodiode made of InAsBi$_{0.02}$ grown by MBE on InAs substrate allowed to demonstrate a long cut-off wavelength of 3.95 µm combined with a lower temperature dependence of the band gap with respect to InAs photodiodes [Sandall-2014].

Regarding the work more specifically focused on the use of Bi with antimonides, a few studies were carried out in order to understand Bi incorporation in GaSb since 2012. The earliest reports on epitaxial GaSbBi alloys show Bi incorporation as low as 0.8% either by MBE or LPE (Fig. 4) [Song-2012, Das-2012]. Attempts to grow GaSbBi alloys by MBE with higher Bi-content resulted in the formation of Sb-Bi droplets on the surface [Duzik-2014]. Varying the growth temperature and the Bi/Sb flux ratio, Rajpalke *et al.* demonstrated GaSbBi epilayers with smooth surface and high crystalline quality. They also observed a reduction of the band-gap of the material corresponding to ~35 meV/%Bi [Rajpalke-2014, Polak-2014]. The PL wavelength shifted up to 3 µm for layers with 9.6% of Bi. In 2017, the University of Montpellier reported a maximum Bi concentration of 14%, reaching a PL emission wavelength of 3.8 µm at RT [Delorme-2017a]. More recently, GaSbBi/GaSb QWs with various thicknesses and compositions were grown [Delorme-2017b, Yue-2018a], leading to the realization of a laser diode emitting near 2.7 µm under pulsed operation at RT. Very preliminary studies were also published on the quaternary alloys AlGaSbBi and GaInSbBi. As expected, the incorporation of Bi resulted in a decrease of the band gap of these alloys, as well as a smaller variation of the band-gap with temperature [Kopaczek-2014]. It was finally found that the incorporation of In into GaSbBi enhanced the PL intensity of the material [Linhart-2017].

In this chapter we review the development of GaSbBi alloys and heterostructures, with an emphasis on recent results.

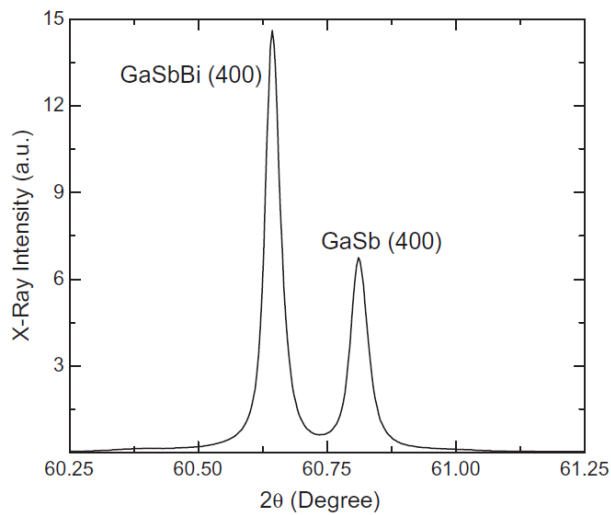

**Fig. 4** HRXRD rocking curve for LPE GaSbBi on GaSb substrate. Reprinted with permission from Elsevier [Das-2012].

## 2. Molecular Beam Epitaxy of GaSbBi

### 2.1. Growth conditions

The first attempts to epitaxially grow GaSbBi were reported in 2012 using LPE [Das-2012] and MBE [Song-2012]. At that time, the Bi content remained very low with both techniques (0.4% and 0.7%, respectively). A better comprehension of the Bi incorporation in GaSb was therefore necessary, and the influence of the growth conditions has been investigated. Since 2012, several groups especially demonstrated that the growth temperature and the V/III flux ratio have a tremendous impact on the Bi incorporation [Rajpalke-2013, Rajpalke-2014, Rajpalke-2015, Delorme-2017a, Yue-2018a]. These parameters need to be carefully adjusted to enhance the Bi incorporation while maintaining a high material quality.

#### 2.1.1. Substrate Temperature

As for other III-V-Bi alloys, it was demonstrated that the Bi incorporation in GaSb requires extremely low growth temperatures. Indeed, due to the higher bonding energy of Ga–Sb than that of Ga–Bi, Bi atoms tend to segregate toward the surface. In 2013, Rajpalke *et al.* managed to increase the Bi content in GaSbBi epilayers from 0.5 to 5% by decreasing the temperature from 350 to 250°C (Fig. 5) [Rajpalke-2013]. High epitaxial quality was obtained, with more than 98% of the Bi atoms in substitutional position. The temperature dependence of the Bi incorporation was fitted using the kinetic model described by Wood *et al.* [Wood-1982] and Pan *et al.* [Pan-2000], previously applied to the incorporation of N in GaInSb alloys [Ashwin-2011]. The Bi content can be expressed using the following equation:

$$\%Bi = \frac{\alpha * J_{Bi}}{\alpha * J_{Ga} + D_0 * \exp\left(-\frac{E_d}{k*T}\right)} \quad (1)$$

where α is a constant. $J_{Ga}$ and $J_{Bi}$ are the Ga and Bi incident flux, respectively. $E_d$ is the energy barrier for Bi desorption (1.75 eV) and $D_0 = 1/\tau_s$, where $\tau_s$ is the surface residence lifetime of the Bi atom (6.5 µs). A similar temperature dependence on the Bi incorporation was also reported in recent publications [Delorme-2017a, Yue-2018a].

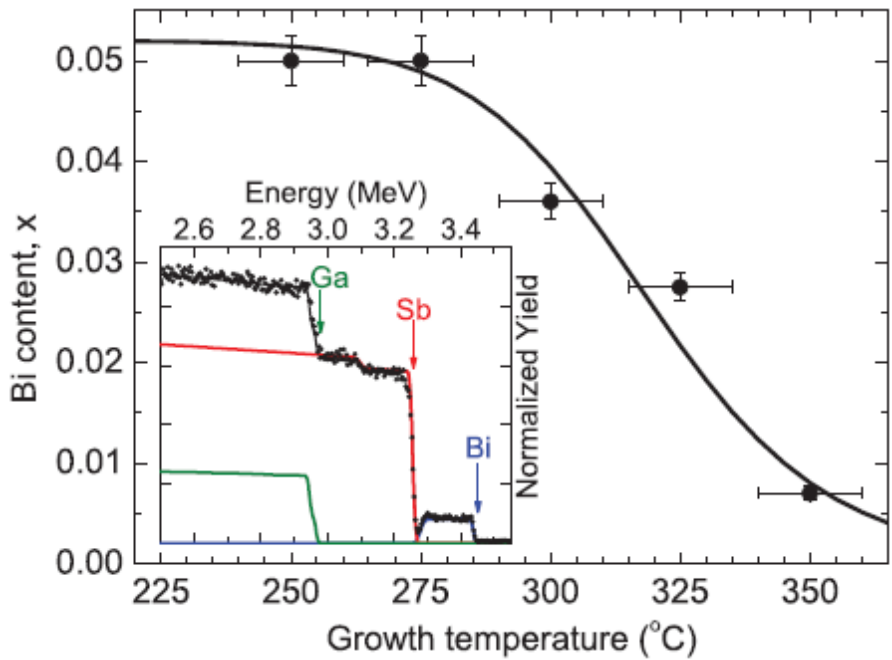

**Fig. 5** Bi content as a function of growth temperature at a fixed growth rate and Bi flux. The points are the experimental data and the solid line is the calculated dependence from the kinetic model. Reprinted with permission from American Institute of Physics [Rajpalke-2013].

As shown in Fig. 5, the optimal growth temperature of GaSbBi is extremely low, about 200°C below the usual growth temperature of others III-Sb alloys. Such low temperatures cannot be measured with conventional pyrometers generally used in III-V MBE systems, suited for temperatures typically higher than 350 to 400°C. The optimization of the growth temperature is therefore challenging.

### 2.1.2. V/III ratio

Due to the low growth temperature of GaSbBi, the Sb desorption from the surface is much weaker than during the epitaxy of other III-V alloys. Therefore, the V/III flux ratio must be close to unity to avoid a metallization of the surface and to maintain a good material quality. It was also recently demonstrated that the Sb flux has a major impact on the Bi incorporation: Delorme *et al.* [Delorme-2017a] and Yue *et al.* [Yue-2018a] reported that an excess of Sb atoms during the growth of GaSbBi causes a sharp drop of the Bi concentration, as shown in Fig. 6. Therefore, the Sb/Ga flux ratio must be kept close to unity to enhance Bi incorporation. However, a lack of group-V elements during growth results in the formation of droplets on the surface and in a degradation of the material quality. A fine tuning of the Sb flux is thus required.

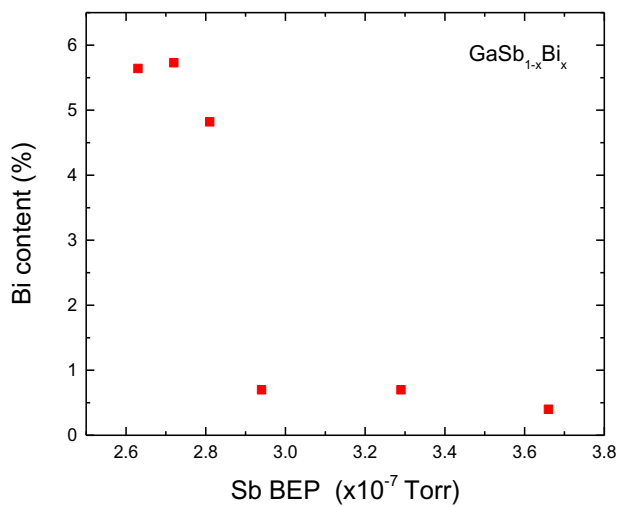

**Fig. 6** The Bi content in GaSb$_{1-x}$Bi$_x$ alloys grown at the same temperature as a function of Sb BEP. The Ga and Bi fluxes were kept constant. Reprinted with permission from Elsevier [Delorme-2017a].

To reach high Bi concentration, the Bi/Sb flux ratio is then a crucial parameter. By increasing the Bi flux, Duzik *et al.* managed to grow GaSbBi epilayers with Bi content up to 12%, but the surface was covered by Ga-Bi droplets [Duzik-2014]. An unintentional As incorporation, increasing with the Bi concentration, was also noticed: for the highest Bi content (12%), an As content as large as 9.3% was estimated by HR-XRD. Rajpalke *et al.* reached a Bi concentration of 9.6% by varying the Bi/Sb flux ratio [Rajpalke-2014, Rajpalke-2015]. Despite the high Bi content, excellent crystal quality was obtained with more than 99% of the Bi atoms in substitutional position. Bi droplets were nonetheless observed on the samples grown under the highest Bi fluxes. More recently, Delorme *et al.* have grown droplet-free GaSbBi layers with a Bi content up to 11.4% by increasing the Bi flux while carefully adjusting the Sb flux to maintain a V/III flux ratio close to unity [Delorme-2017a]. Well-defined *Pendellösung* fringes on the HR-XRD scans indicated smooth interfaces and homogeneous composition. For higher Bi concentrations, Ga-Bi droplets were systematically observed on the surface and damped *Pendellösung* fringes were measured, showing a degradation of the crystal quality (Fig. 7). Finally, a maximum Bi concentration of 14% was reached.

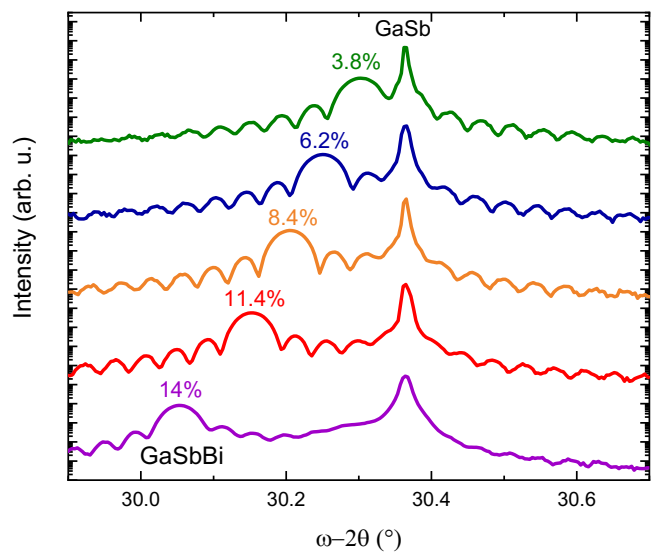

**Fig. 7** HRXRD scans of the 004 Bragg reflection of Bi flux dependent GaSbBi films with Bi content of 3.8, 6.2, 8.4, 11.4% and 14% grown at the same temperature and growth rate. The Sb flux was adjusted to maintain a V/III flux ratio close to stoichiometry. All these samples are droplet free, except the one with 14% Bi. Partly reprinted with permission from Elsevier [Delorme-2017a].

## 2.2. Setting the growth parameters

Overall, the conditions (temperature and flux ratio) allowing the growth of high quality GaSbBi are extremely challenging to set since any slight deviation leads either to the formation of droplets or to a low incorporation of Bi into GaSb. The usual way to set the growth conditions is to grow different sets of samples to find the optimal temperature for the Bi incorporation, to adjust the Ga and Bi flux to get the desired growth rate and Bi content, and finally to fine tune the Sb flux in order to obtain the optimum V/III ratio. A large number of calibration samples is thus generally required to correctly set these parameters by a trial-and-error approach, which is both time- and money- consuming.

Recently, Delorme *et al.* reported a new method to set the optimized growth conditions of GaSbBi alloys, based on the RHEED intensity oscillations [Delorme-2018]. To demonstrate the possibility to use RHEED oscillations for adjusting the Sb flux, three series of measurements with three different Bi fluxes corresponding respectively to Bi contents of approximately 4, 8, and 12% were carried out. The Sb flux was varied while the Ga incorporation rate and the temperature were kept constant for each set of experiments. Despite the extremely low growth temperatures, strong oscillations intensity were observed for both GaSbBi and GaSb (Fig. 8 a and b). For an Sb/Ga ratio above unity, the GaSbBi and GaSb growth rates calculated from the RHEED oscillations were similar (Fig. 8 c, d and e), confirming that the Ga incorporation rate is the limiting mechanism. For an Sb/Ga ratio slightly below unity, damped oscillations were recorded during GaSb growth. As expected, the GaSb growth rate decreased significantly, clearly indicating an Sb limited growth. The oscillations recorded at the same Sb beam equivalent pressure (BEP) during GaSbBi growth were comparable to the ones obtained at higher Sb flux, revealing a (Sb+Bi)/Ga flux ratio larger or very close to 1. The difference between the growth rates of GaSbBi and GaSb was attributed to the Bi incorporation. Following each set of RHEED oscillations measurements, 100-nm thick GaSb$_{1-x}$Bi$_x$ epilayers with 0 < x < 13% were grown using various Sb/Ga and Bi/Sb ratios to confirm the possibility to use these data to accurately adjust the Sb flux to the chosen Ga rate and Bi content. Overall, the samples with the highest Bi incorporation and the best material quality were obtained for a (Sb+Bi)/Ga ratio slightly higher than unity (< 1.1).

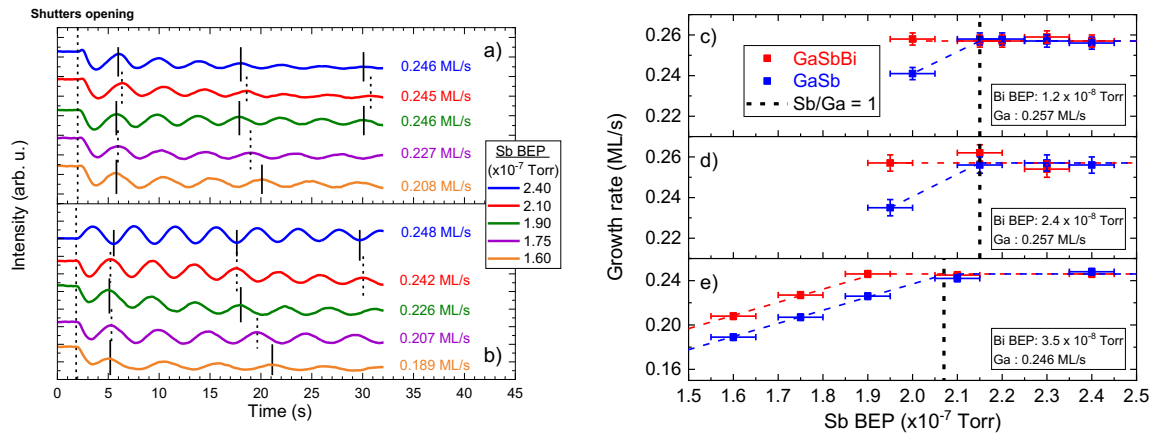

**Fig. 8** RHEED oscillations recorded for GaSbBi (a) and GaSb (b) grown as a function of the Sb flux. The temperature, Bi and Ga fluxes were fixed. The curves have been vertically shifted for clarity, and black lines are a guide for the eye. (c), (d), (e) GaSbBi and GaSb calculated growth rates as a function of the Sb flux using different Bi BEPs. Each data point is the average of several measurements. Reprinted with permission from Elsevier [Delorme-2018].

Delorme *et al.* also demonstrated that the other critical parameter for the Bi incorporation, the growth temperature, could be set using RHEED oscillations. As mentioned above, the growth rate of GaSbBi and GaSb is different below or very close to the Ga-Sb stoichiometry, due to the incorporation of Bi. Thus, by setting the Sb BEP at a value where the GaSbBi growth rate is limited by the Ga incorporation while it is limited by the Sb incorporation for GaSb, it is possible to clearly measure the decrease of the Bi incorporation induced by the increase of the substrate temperature. The GaSbBi growth rate was measured between 170 and 320°C (thermocouple temperature reading (TTR)), revealing the expected temperature dependence of the growth rate (Fig. 9a). This temperature dependence corresponds to the variation of the Bi incorporation and can thus be modeled by the kinetic approach presented above (Fig. 5) and using the same parameters ($E_d$ and $\tau_s$), as confirmed by Fig. 9b.

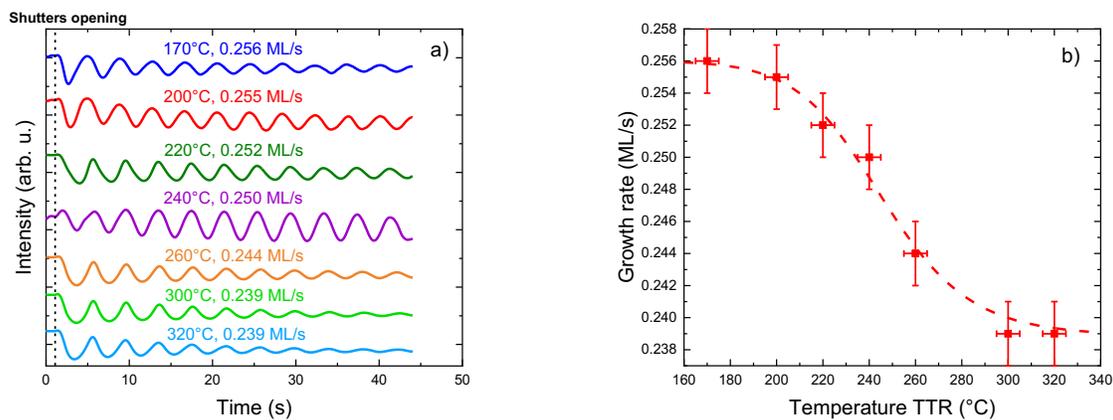

**Fig. 9** (a) RHEED oscillations from GaSbBi grown at different temperatures using fixed fluxes. The calculated growth rates are indicated for each curve. The curves have been vertically shifted for clarity. (b) GaSbBi growth rate calculated from the RHEED oscillations as a function of the growth temperature. The dashed line was calculated using a kinetic model. Reprinted with permission from Elsevier [Delorme-2018].

# 3. Microstructure of Ga(Sb,Bi) investigated by Transmission Electron Microscopy (TEM)

## 3.1. Introduction

III-V-Bi compounds are highly-mismatched alloys (HMA) which are formed by the isoelectronic substitution of elements with very different size and/or electronegativity in the anion sublattice. As a consequence, HMA are often affected by miscibility gaps, which makes their growth challenging due to the phase separation tendency of the alloy. Extensive work on other HMA systems (*e.g.* dilute group-III nitrides) evidenced that composition fluctuations and morphological instabilities have a detrimental effect on the optical response and limits further optoelectronic applications [Trampert-2004]. Among Bi-containing III-V semiconductors, Ga(As,Bi) is the most investigated material. Recently, Gibbs formation energies and the immiscibility curve of Ga(As,Bi) were theoretically determined using different methods, evidencing that the thermodynamics of $GaAs_{1-x}Bi_x$ alloys strongly favor decomposition [Punkkinen-2015]. As a matter of fact, experimental studies have demonstrated that under specific circumstances cluster formation, phase separation and atomic ordering do occur in Ga(As,Bi) epilayers [Norman-2011, Sales-

2011, Bastiman-2012, Butkuté-2012, Puustinen-2013, Duzik-2014, Reyes-2014, Wu-2014a, Wu-2014b, Lu-2015, Luna-2015, Luna-2016, Beyer-2017, Butkuté-2017, Wood-2017, Balades-2018].

Compared to other III-V-Bi compounds, GaSb compounds alloyed with Bi have been far less studied (the first reports on the alloy date back to 2012 [Das-2012, Song-2012]) and very basic material-related questions such as alloy stability, segregation or solubility limits remain completely unexplored. Although the mismatch in atomic radius and electronegativity between Sb and Bi is smaller than between As and Bi and thus a higher solubility of Bi into GaSb than that of Bi into GaAs is expected, the incorporation mechanisms and solubility limits of Bi into GaSb are still unknown. Indeed, the maximum Bi incorporated in GaSb to date is about 14% as determined using Rutherford backscattering spectroscopy (RBS) [Delorme-2017a]. This is smaller than the reported maximum 22% Bi incorporation into GaAs [Lewis-2012], as determined by HR-XRD and assuming Vegard's law and a GaBi lattice constant of 6.33 Å, extrapolated from RBS measurements [Tixier-2003].

At present, most of the few published works on Ga(Sb,Bi) focus on the growth by MBE of the material to explore the growth parameters controlling Bi incorporation [Song-2012, Rajpalke-2013, Rajpalke-2014, Rajpalke-2015, Delorme-2017a, 2018]. The assessment of the quality of the samples is commonly based on XRD, atomic force microscopy, scanning electron microscopy and optical spectroscopy measurements. The Bi composition is mainly determined using RBS [Song-2012, Rajpalke-2013, Rajpalke-2014, Duzik-2014, Rajpalke-2015]. Surprisingly, until very recently (2017-2018) there was a lack of information on the samples' microstructure determined using TEM, an otherwise powerful tool to get experimental evidence of the structural quality of the layers, interface abruptness and estimations of the local chemical composition.

### 3.2. Experimental details

One of the reasons for the apparent lack of works on TEM investigations of Ga(Sb,Bi) may originate from the fact that the study poses several difficulties. One of the very first limitations concerns the delicate TEM specimen preparation on its own, which requires the development of specific preparation techniques/steps to assure high-quality Ga(Sb,Bi) thin foils free of artefacts introduced during preparation. The other important limitation concerns the low contrast/low signal of Ga(Sb,Bi) in (S)TEM measurements, as will be discussed next. Note that due to the low contrast and, in general, the low signal of Ga(Sb,Bi) at (S)TEM, the measurements are very sensitive to strain, thin-foil surface relaxation and bending, hence the successful (S)TEM investigation of Ga(Sb,Bi) demands extremely high quality specimens.

### 3.2.1. Sample preparation

To date, the scarce TEM investigations of Ga(Sb,Bi) refers to cross-sectional observations where the samples are commonly prepared in the [110] and [1̅10] projections. Although the mechanical properties of Ga(Sb,Bi) are unknown, first TEM works report that Ga(Sb,Bi) is a very soft material, even softer than other GaSb-based related compounds. It gets thus very easily damaged during TEM preparation [Luna-2018b]. Furthermore, due to the expected metastable character of the HMA Ga(Sb,Bi), TEM specimen preparation may modify the pristine structure and introduce artifacts. In particular, the threshold in temperature and/or energy (Ar-/Ga-ions) to induce morphological changes and/or cluster formation in Ga(Sb,Bi) is unknown. In short, there are two widely used methods to prepare TEM specimens, (i) the conventional preparation including mechanical grinding, dimpling, followed by final Ar-ion milling, and (ii) focus ion beam (FIB).

The most critical steps in conventional TEM preparation of Ga(Sb,Bi) concern the dimple grinding and Ar-ion milling processes. In order to minimize the damage introduced during dimpling, a low speed rotation is recommended, together with the use of diamond paste with 1 and ¼ µm grain size. In addition, Ga(Sb,Bi) TEM specimens are extremely sensitive to the damage introduced during ion milling at precision ion polishing system (PIPS$^{TM}$ Gatan, Inc.). It has been recently reported that the best quality specimens are obtained while sputtering with very low energy Ar-ions, in the range of 2–0.5 keV [Luna-2018b]. The use of a cooling stage at PIPS seems to be less critical than the low energy of the Ar beam. Fig. 10 illustrates some of above mentioned difficulties. The images in Fig. 10 a and b correspond to early Ga(Sb,Bi) TEM specimens which were prepared following the standard procedure for GaSb-based compounds (without Bi). Note the damage introduced during PIPS, with strong material redeposition (Fig. 10a) and preferential sputtering (Fig. 10b). A similar damage is reported for specimens prepared by FIB (Fig. 10c), with stripes across the whole structure introduced during the specimen preparation process [Yue-2018a]. On the other hand, Fig. 10 d and e show images from specimens prepared following the optimized steps in conventional preparation described above. The improvement is noticeable and although it is obvious that further improvements are still required, the quality of the specimens is sufficient to allow TEM investigations and a first insight into the microstructure and chemical properties of Ga(Sb,Bi) epilayers and QWs.

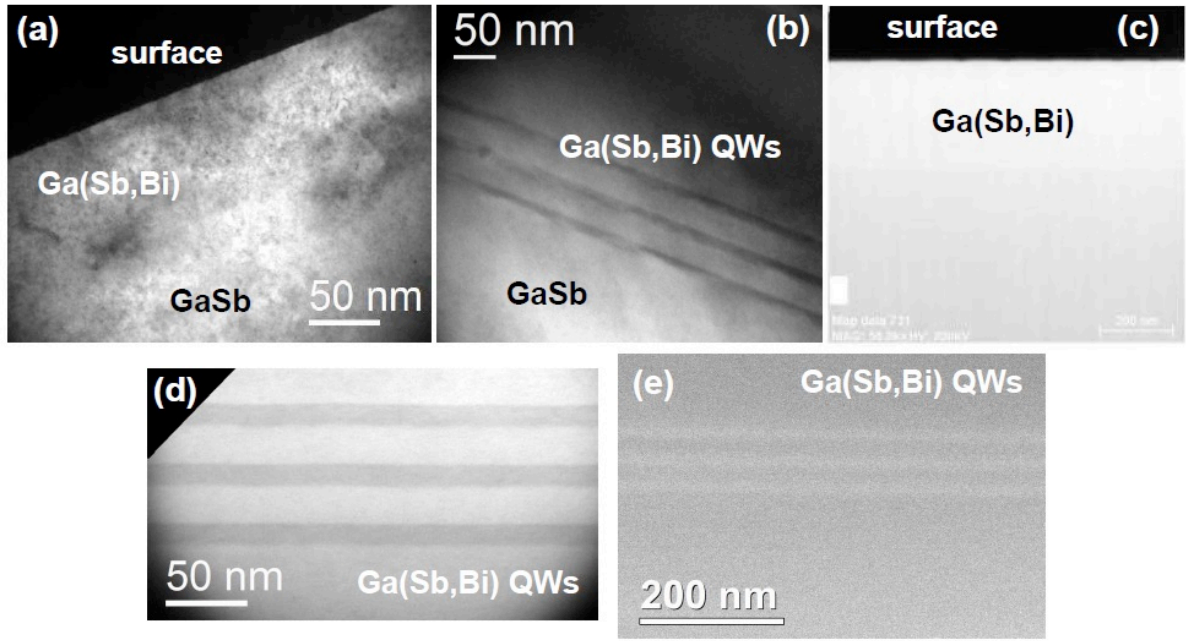

**Fig. 10** Cross-sectional TEM specimen preparation: Material redeposition (a) and preferential sputtering (b) introduced during ion milling at PIPS in Ga(Sb,Bi)-based TEM specimens prepared using the standard method and (c) stripes originating during ion milling in FIB TEM specimen preparation. (d)-(e) Bright-field and HAADF micrographs of the same QWs in (b) obtained using conventional TEM specimen preparation after improving the critical steps of dimpling and ion-milling at PIPS. (c) is reprinted with permission from Elsevier [Yue-2018a].

### 3.2.2. Challenges in the determination of the Bi content in Ga(Sb,Bi)

In addition to the microstructural investigation of Ga(Sb,Bi) epilayers and QWs (for instance to determine and/or rule out the presence of defects), one important aspect of any (S)TEM investigation concerns its analytical capabilities, which would enable the determination of the chemical composition at the atomic scale. This kind of investigation is particularly important in Ga(Sb,Bi) due to its assumed HMA character and the tentative formation of composition modulations and/or clusters, as it is found in other dilute bismide compounds. The mismatch in atomic radius and electronegativity between Sb and Bi is smaller than *e.g.* between As and Bi, which implies that whether Ga(Sb,Bi) can be even referred to as HMA, and behaves as such, is under current investigation.

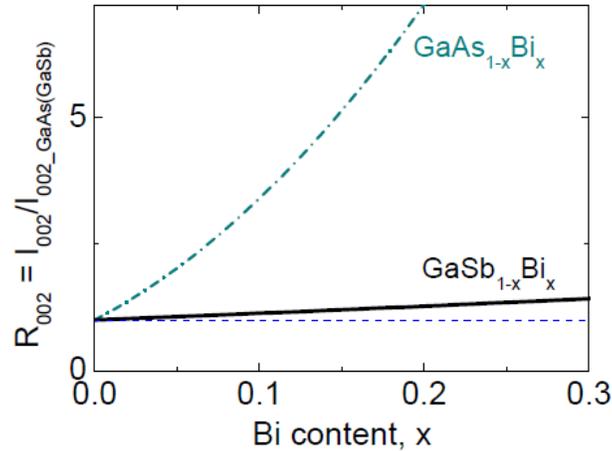

**Fig. 11** Estimated $g_{002}$ DFTEM image contrast computed as the ratio of the diffracted intensity at the layer and at the GaSb (GaAs) reference. Reprinted with permission from the American Institute of Physics [Luna-2018a].

In general, $g_{002}$ dark-field (DF)TEM is a powerful and direct method to determine the element distribution in III-V semiconductors with ZB structure [Bithell-1989, Luna-2009, Lu-2016]. The reason is that in III–V alloys with ZB structure, the diffracted intensity for the 002 reflection under kinematic approximation is proportional to the square of the structure factor, which in turn depends on the difference in the atomic scattering factors of the alloy components. Therefore, in this "structure-factor imaging mode", the contrast mainly arises from differences in the atomic–scattering factors between the group-III and group-V elements. It is shown that $g_{002}$ DFTEM imaging is a very valuable technique for the investigation of Ga(As,Bi) [Luna-2015, Luna-2016, Patil-2017] since it allows the detection of composition fluctuations which cannot be detected using conventional XRD techniques [Wu-2015]. Theoretical estimations of the diffracted intensity for the 002 reflection in Ga(Sb,Bi) predict, however, that the intensity contrast to GaSb is extremely low [$I_{002-GaSbBi}/I_{002-GaSb}$ ~ 1.13, for Ga(Sb,Bi) with 14% Bi], which is in marked difference to Ga(As,Bi) [$I_{002-GaAsBi}/I_{002-GaAs}$ ~ 4.7 for Ga(As,Bi) with 14% Bi], where small amounts of incorporated Bi result in a strong image contrast as illustrated in Fig. 11. A reliable method to determine the Bi content from $g_{002}$ DFTEM is to follow the procedure proposed by Bithell and Stobbs [Bithell-1989], based on the analysis of the intensity ratio $R_{002} = I_{002}/I_{002}^{GaSb}$. Assuming the kinematical scattering approximation (with atomic-scattering factors adapted from Doyle and Turner [Doyle-1968]) and the validity of Vegard´s law, the following dependence of $R_{002}$ on the Bi content x in GaSb$_{1-x}$Bi$_x$ is obtained: $R_{002} = (1 + 0.638 x)^2$, which is plotted in Fig. 11 for x ranging between 0 and 0.3.

Due to the low contrast of $g_{002}$ DFTEM, this technique demands extremely high quality TEM specimens with smooth surfaces and reduced buckling, since this requirement is crucial to assure accurate $g_{002}$ diffraction conditions. Hence, in order to estimate the Bi content and, in particular, its spatial distribution (*i.e.* homogeneity), it is recommended to correlate results from $g_{002}$ DFTEM with high-angle annular dark-field (HAADF) measurements, local energy dispersive x-ray spectrometry (EDS) measurements, electron energy-loss spectroscopy (EELS), as well as macroscopic and highly-localized XRD studies. Note that EELS and EDS data in Ga(Sb,Bi) are predicted to be intrinsically difficult to interpret because of the close atomic numbers of the constituents which result in overlapping spectral features. Furthermore, the lack of suitable references (*e.g.*, the not-yet synthesized endpoint GaBi compound) also limits EDS capabilities for chemical quantification. Although HAADF measurements with so-called Z-contrast exhibits a weaker contrast than expected due to the close proximity of $Z_{Sb} = 51$ and $Z_{Bi} = 83$ ($I_{HAADF} \propto Z^{1.7}$) as shown in Fig. 10e and Fig. 12, and hence, depending on the amount of Bi incorporated in the sample, the technique may not be sufficiently informative to enable estimation of the Bi content, HAADF images seem nevertheless to provide relevant information on the material homogeneity. The TEM specimens used for HAADF in Fig. 12 exhibit noticeable thickness variations, since the measurements are performed at relatively thick areas to avoid the strong damage by PIPS in the thinner STEM suitable areas.

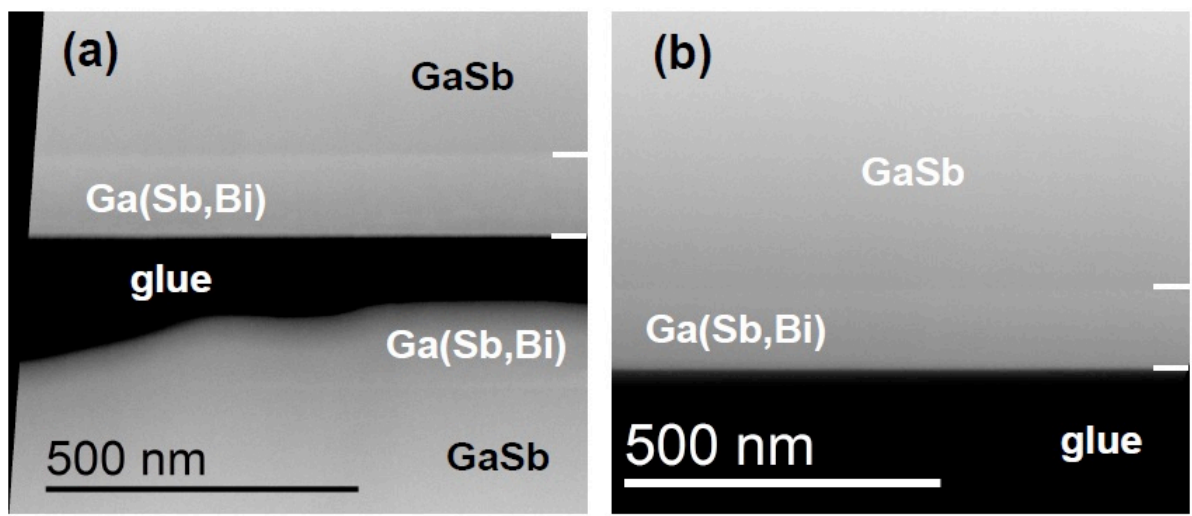

**Fig. 12** HAADF overview images of Ga(Sb,Bi) epilayers with about 14% Bi (a) and 11%–12% Bi (b). Reprinted with permission from the American Institute of Physics [Luna-2018a], changed layout.

### 3.3. Ga(Sb,Bi) epilayers

MBE-grown Ga(Sb,Bi) epilayers on GaSb(001) substrates with up to 14% Bi have been investigated by TEM [Luna-2018a]. The epilayers grow pseudomorphically up to 120 nm thickness and no clusters, dislocations or extended defects have been detected (cf. Fig. 12 and 13) In addition, Yue *et al.* have reported the growth of 280 nm thick high Bi content (up to 11%, locally 13%) Ga(Sb,Bi) epilayers. Assuming a lattice constant of $a_{GaBi}$= 6.3 Å [Yue-2018a] (the exact value of $a_{GaBi}$ is under discussion [Tiedje-2008, Rajpalke-2014, Wang-2017, Delorme-2017a Yue-2018a]) and using the elastic constants of GaSb [the mechanical properties and elastic constants of Ga(Sb,Bi) are unknown], these layers are thinner than the critical thickness calculated with Matthews and Blakeslee model [Matthews-1974]:

$$h_c = \frac{b}{2\pi f} \frac{(1-\nu \cos^2 \alpha)}{(1+\nu)\cos \lambda} \left( \ln \frac{h_c}{b} + 1 \right) \qquad (2)$$

where b is the dislocation Burgers vector, $\nu$ is the Poisson ratio, $\alpha$ is the angle between the dislocation line and its Burgers vector and $\lambda$ is the angle between the slip direction and that direction in the film plane which is perpendicular to the line of intersection of the slip plane and the interface [Matthews-1974]. In the case of 280 nm thick Ga(Sb,Bi) epilayers with 5.6% Bi [Yue-2018a], $f$ ~ 0.19% and $h_c$ ~ 404 nm. Thus, the above mentioned layers would still be below the estimated limit for plastic relaxation, which explains the absence of misfit dislocations in these cases. As a matter of comparison, note for instance that the lattice mismatch of $In_{0.2}Ga_{0.8}As$ to GaAs(001) is $f$ ~ 1.4%, which results in a critical thickness of $h_c$ ~ 34 nm.

Figure 13a displays a $g_{002}$ DFTEM micrograph of a Ga(Sb,Bi) epilayer with 14% Bi, the maximum Bi incorporated into GaSb so far. Local quantitative chemical determination from the analysis of the $g_{002}$ diffracted intensity yields an average Bi content [Bi] = (14.2 ± 0.8)%, which is in good agreement with $[Bi]_{RBS}$ = 14%. The error bar refers to the standard deviation of the data. On the other hand, the low contrast in $g_{002}$ DFTEM imaging of Ga(Sb,Bi) renders the detection of CMs an arduous task. It is reported that the amplitude of lateral CMs in Ga(As,Bi) is about $\Delta[Bi]/[Bi]$ = 30% [Luna-2015, Wu-2015, Luna-2016]. In Ga(As,Bi) with an average Bi content of 5%, the presence of a 30% modulation induces a contrast of about 16%, which is readily detectable since the estimated lower limit of detection using $g_{002}$ DFTEM is around 3–3.6% contrast [Luna-2017]. Regarding Ga(Sb,Bi), little is known about the existence, character and/or magnitude of CMs in this material system. Assuming the amplitude of CMs in Ga(Sb,Bi) is as large as that in Ga(As,Bi), then a 30% modulation in Ga(Sb,Bi) with an average Bi content of 14% would induce a contrast of about 5.7%, which is significantly smaller than in Ga(As,Bi) but still detectable. Yet a modulation of 20% in Ga(Sb,Bi) with 14% Bi would be at the limit of detection, with a 3.4% contrast. EDS measurements of the local composition can provide complementary information to $g_{002}$ DFTEM, as shown in Figure 13b where EDS maps of the chemical composition also evidence a homogeneous layer in agreement with the $g_{002}$ DFTEM contrast in Fig. 13a. HAADF micrographs (cf. Fig. 12) provide similar information to $g_{002}$ DFTEM and EDS: the layers are homogeneous with CMs below 20%.

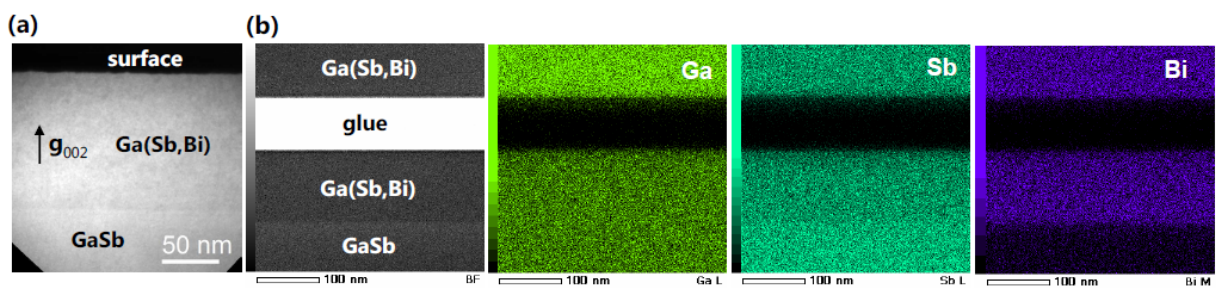

**Fig. 13** (a) Chemically sensitive $g_{002}$ DFTEM micrograph and (b) bright-field STEM image and EDS compositional map of a pseudomorphic Ga(Sb,Bi) epilayer on GaSb(001) with 14% Bi. (b) is reprinted with permission from the American Institute of Physics [Luna-2018a].

On the other hand, the reported Ga(Sb,Bi) epilayers are free of extended defects and/or dislocations, but nevertheless there are surface droplets, in particular for higher Bi containing layers, *i.e.* with Bi > 11-12% [Duzik-2014, Delorme-2017a, Yue-2018a, Luna-2018a]. In layers with Bi contents larger than 11-12%, surface irregularities are also observed (cf. Fig. 14 a and b). Surface droplets are pure Bi droplets and/or Bi/Ga droplets, *i.e.* the droplets are composed of two separated phases, as illustrated in the micrograph in Fig. 14c [Duzik-2014,Yue-2018a]. Although the specific impact of surface droplets is under investigation, recent works indicate that Bi droplets seem to drastically affect Bi incorporation in Ga(Sb,Bi) leading to inhomogeneous chemical distributions [Duzik-2014,Yue-2018a]. Thus, their impact on Ga(Sb,Bi) would be similar to that reported for Ga(As,Bi) [Steele-2016,Wood-2016,Tait-2017]. Examples of such inhomogeneities in Ga(Sb,Bi) are the observation of a "two-composition" layer [Luna-2018a], cf. Fig. 14b, or sequences of several irregular layers (in shape and thickness) with distinct Bi contents [Tait-2016, Yue-2018a]. For instance, Yue *et al.* reported on a 280 nm-thick-film of Ga(Sb,Bi) consisting of a ~60 nm layer close to the substrate with 5-7.5% Bi, followed by a 170 nm intermediate region of about 1% Bi and with a 22-50 nm top layer close to the surface with the highest 13% Bi [Yue-2018a].This unintentional "multi-layer" sequence has been already observed in Ga(As,Bi) [Norman-2011,Reyes-2014,Wood-2016,Steel-2016,Tait-2017] and is linked to Bi surface segregation leading to Bi accumulation and droplet formation. Several models have been proposed to explain the droplets formation and Bi incorporation under these conditions [Tait-2016, Wood-2016, Steel-2016]. Furthermore, in addition to the inhomogeneous Bi incorporation, Duzik *et al.* and Luna *et al.* have found evidences of droplet etching into the underlying film [Duzik-2014, Luna-2018a], an example can be seen in Figs. 12a and 14 a and b.

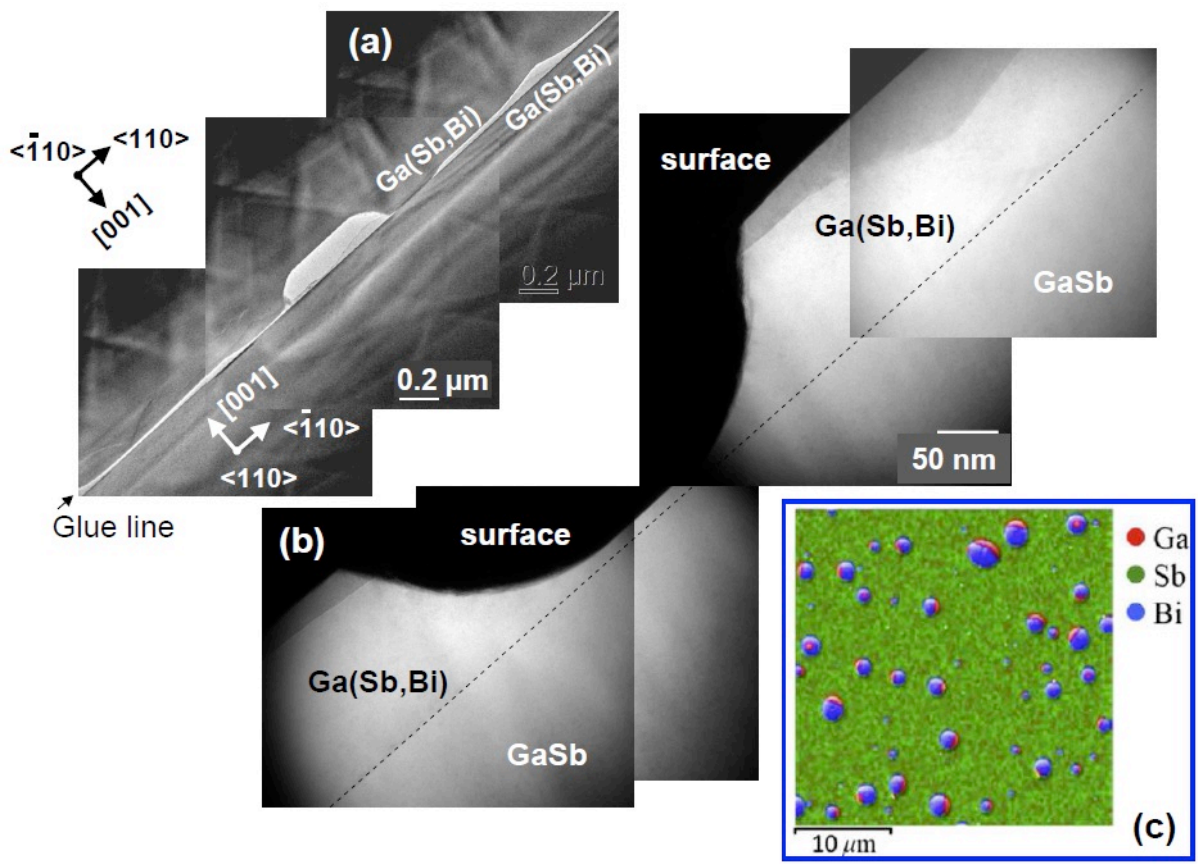

**Fig. 14** (a)-(b) Overview TEM micrographs displaying assumed etching features at the surface of Ga(Sb,Bi) epilayers, which seem to originate from (Bi) surface droplets. (c) EDS mapping of the droplets in the surface of a Ga(Sb,Bi) epilayer. (c) is reprinted with permission from Elsevier [Yue-2018a].

In summary, the presence of droplets may have locally dramatic consequences on the microstructure of the underlying film. Nevertheless, in areas without surface irregularities, the Ga(Sb,Bi) layers are rather homogeneous in composition with CMs, if present at all, below 20%.

## 3.4. Ga(Sb,Bi)/GaSb quantum wells

Most TEM investigations focus on the analysis of Ga(Sb,Bi)/GaSb QWs. As the active zone of future mid-IR laser/detector structures [Delorme-2017b], QWs with reproducible and well-controlled composition and QW/barrier thicknesses are a prerequisite.

### 3.4.1. Interface analysis

As a matter of fact, the high structural quality of the dilute bismide material also reflects in Ga(Sb,Bi)/GaSb QWs. The QWs displayed in Fig. 15a and b correspond to 7 and 15 nm Ga(Sb,Bi)/GaSb QWs, respectively, with 11% Bi. The thickness of the QWs is well below the estimated critical thickness for the introduction of dislocations, $h_c \sim 185$ nm, which explains the absence of strain-relieving defects, such as dislocations.

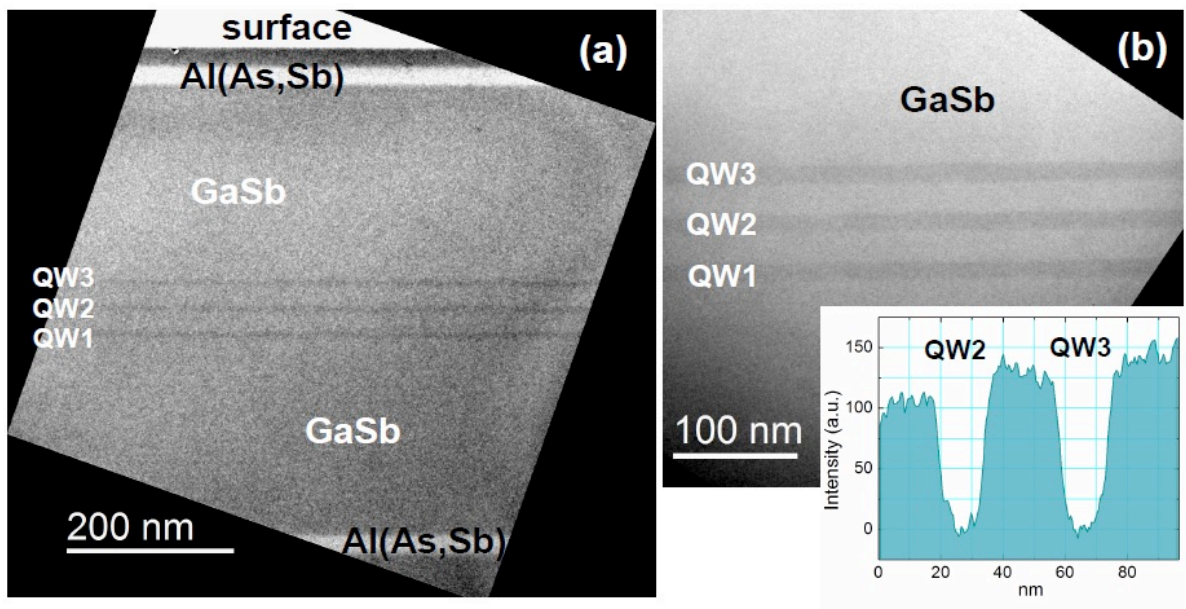

**Fig. 15** Overview bright-field TEM images of (a) 7 nm and (b) 15 nm Ga(Sb,Bi)/GaSb QWs with 11% Bi.

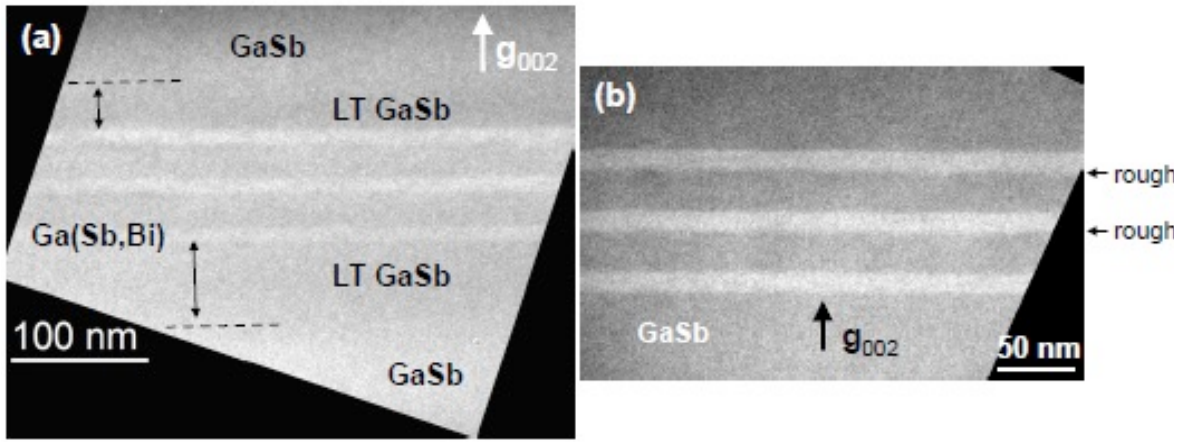

**Fig. 16** Representative chemically sensitive $g_{002}$ DFTEM micrographs of Ga(Sb,Bi)/GaSb QWs, displaying lateral thickness fluctuations and (a) a faint change in contrast between the GaSb-layer and the LT grown GaSb, which could be tentatively attributed to the presence of point defects in the LT GaSb barriers and (b) QW thickness variations and the smoothing effect at the upper GaSb-on-Ga(Sb,Bi) interface. From [Luna-2018b].

Fig. 16a displays a chemically sensitive cross-section $g_{002}$ DFTEM micrograph of a representative sample with nominally 11 nm thick QWs with 11% Bi. As observed, the Ga(Sb,Bi) QWs exhibit regular and homogeneous morphologies including smooth interfaces. In the image it is possible to identify a faint change in contrast between the GaSb-layer and the low-temperature (LT) grown GaSb, which could be tentatively attributed to the sensitivity of $g_{002}$ DFTEM to even detect local variations in point defect density, as reported by Glas *et al.* [Glas-2004]. Hence, this contrast change could be an indication of the presence of point defects in the LT GaSb barriers, which would likely have an impact on the emission properties of the Ga(Sb,Bi)/GaSb QW structures.

On the other hand, despite the QWs look at first glance very regular and homogeneous in composition (cf. Fig. 16), a detailed examination reveals that the three QWs are affected by lateral thickness fluctuations (on the 100-nm-length scale of the TEM images) and exhibit a considerable interface roughness, in particular at the lower Ga(Sb,Bi)-on-GaSb interface.

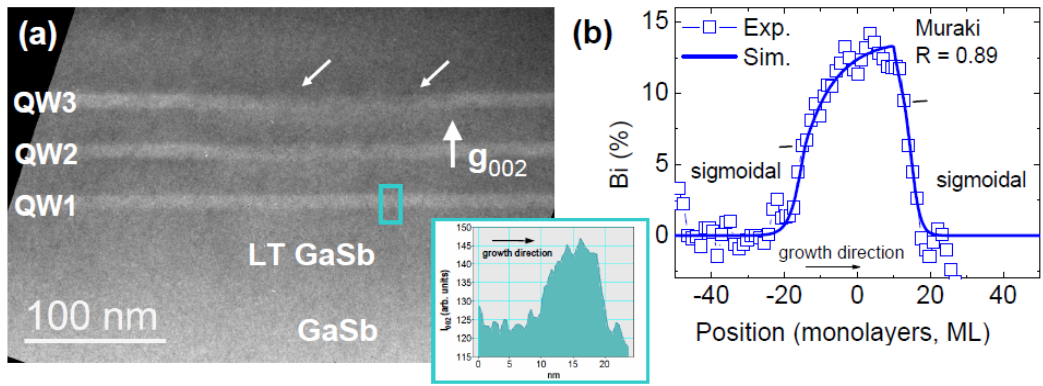

**Fig. 17** (a) Chemically sensitive g002 DFTEM micrograph of Ga(Sb,Bi)/GaSb QWs together with (b) a representative Bi composition profile across the QW, extracted from the analysis of the **g**$_{002}$ DFTEM diffracted intensity. The experimental Bi distribution shows clear signatures of Bi segregation. The experimental data are fitted using a combination of sigmoidal functions to account for the intrinsic interface broadening and of Muraki's segregation model, from where the segregation efficiency R is extracted. Note the local irregularities in the growth of QW3 and marked by white small arrows. From [Luna-2018b].

A significant observation is the fact that the upper GaSb-on-Ga(Sb,Bi) interface is smoother than the lower Ga(Sb,Bi)-on-GaSb interface. This observed roughness does not seem to relate to the Bi composition (for the range of Bi contents 4-11%) since different samples exhibit a similar morphology. Hence, the effect does not arise from the accumulation of epitaxial strain due to the increased Bi content or relates to the development of morphological instabilities due to local composition fluctuations [Trampert-2004, Volz-2005]. In those cases, the roughening would predominantly affect the upper GaSb-on-Ga(Sb,Bi) interface, contrary to the experimental observations. A similar interface smoothing at the upper GaSb-on-Ga(Sb,Bi) interface of 10% Bi, 6 nm thick Ga(Sb,Bi)/GaSb QWs has been reported by Yue *et al.* [Yue-2018b] and ascribed to the surfactant effect of Bi. Bi would then alleviate the roughness introduced due to the LT-growth, in particular at the GaSb barriers. Furthermore, a connection between the surfactant effect and Bi segregation is anticipated, as will be discussed later. On the other hand, a clear variation in QW thickness among the three QWs is detected in Fig. 16b. Additionally, the QWs are affected by lateral thickness fluctuations. The impact of the fluctuations is particularly noticeable in the samples with the smallest QW thickness of 6-7 nm, since in this case their amount is on the same order as the QW thickness.

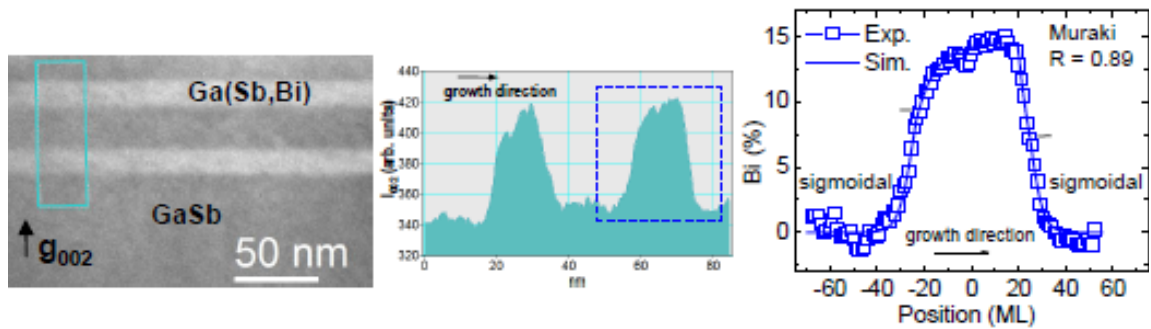

**Fig. 18** Chemically sensitive **g**$_{002}$ DFTEM micrograph of representative Ga(Sb,Bi)/GaSb QWs together with the Bi composition profile across the QW, extracted from the analysis of the **g**$_{002}$ DFTEM diffracted intensity. The experimental data are fitted using a combination of sigmoidal functions to account for the intrinsic interface broadening and of Muraki's segregation model, from where the segregation efficiency R is extracted.

### 3.4.2. Interface undulations and Bi segregation

Besides, in some samples, the QWs are affected by local morphological irregularities as those marked by a small white arrow in Fig. 17a. The origin of these features, mainly affecting the third QW, might relate to Bi segregation and its tendency to accumulate forming droplets, as was discussed in the context of Ga(Sb,Bi) epilayers. Although in this particular case there were no direct evidence of the presence of (surface) droplets in the sample, the lateral extent of the features (about a hundred nm) and its good match with the dimensions of the small surface droplets reported by Yue *et al.* [Yue-2018a] and displayed in Fig. 14c, suggest that these local irregularities may be caused by the presence of Bi droplets accumulated at the interface and further evaporated during growth. Remarkably, Bi distribution profiles across the QW obtained from the analysis of the **g**$_{002}$ DFTEM diffracted intensity [Luna-2018a] reveal clear signatures of Bi surface segregation as shown in the profile in Fig. 17b. As it was previously determined for other III-V heterostructures exhibiting surface segregation [Luna-2008, Lu-2016], the experimental profile in Fig. 17b is very well described by a combination of the intrinsic interface broadening based on sigmoidal functions [Luna-2012] and of Muraki´s phenomenological segregation model [Muraki-1992]**.** In the present case, the segregation efficiency, R, which defines the fraction of Bi atoms in the topmost layer that segregates into the next layer, is about R = 0.89 which is significantly smaller than the value R = 0.9–0.96 reported for the Ga(As,Bi)/GaAs QWs in Ref. [Luna-2016] but larger than R = 0.79 for the Ga(As,Bi)/GaAs QWs grown using a two-substrate-temperature technique by Patil *et al.* [Patil-2017]. It is important to note that these Ga(Sb,Bi)/GaSb QWs exhibit a clear smoothing at the upper GaSb-on-Ga(Sb,Bi) interface, as displayed in Fig. 17 and 18.

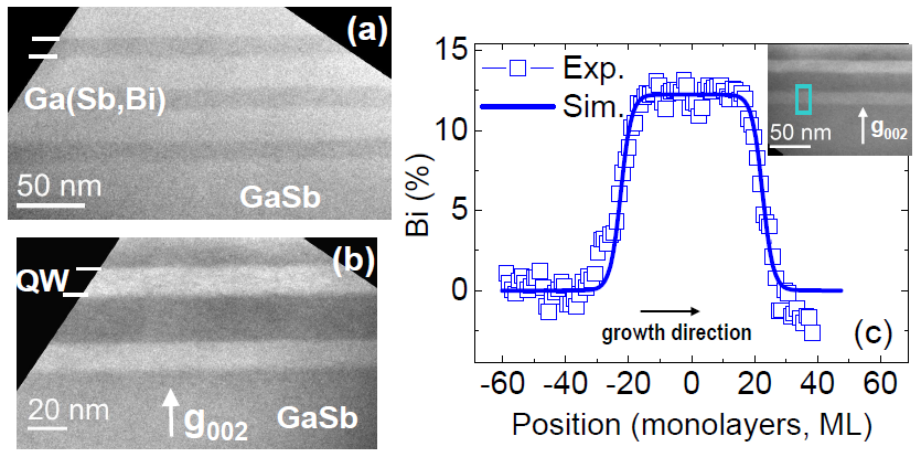

**Fig. 19** (a) Bright-field and (b) chemically sensitive g002 DFTEM micrographs of the Ga(Sb,Bi)/GaSb QWs in a reference (laser) structure, together with (c) the Bi composition profile extracted from the analysis of the $g_{002}$ DFTEM diffracted intensity in the area marked in the inset. The experimental data are fitted to a sigmoidal function. Partly reprinted with permission from the American Institute of Physics [Luna-2018a].

On the other hand, the QWs comprising the reference (laser) structure and the laser itself (see next section) were characterized by narrower and symmetric interfaces without traces of Bi segregation, as shown in Fig. 19 and, in particular, in the experimental distribution profile in Fig. 19c. In this case, the perfect fit to a sigmoidal function further allows quantification of the chemical interface width ranging between 2.2 and 2.7 nm (defined by 10–90% criterion). The estimated width is on the same order as the chemical interface in other III–V heterointerfaces, *e.g.*, 2.1 nm for high quality (Al,Ga)As/GaAs [Luna-2012].

Hence, the presence of Bi segregation can be linked to the Bi surfactant effect and the smoothing of the upper GaSb-on-Ga(Sb,Bi) interface. Such a smoothing effect was barely detectable in the reference (laser) sample (cf. Fig. 19) and in the laser structure where both (narrow) interfaces exhibited a similar roughening.

## 3.5. Laser

Figure 20 displays an overview of the laser structure with an enlarged micrograph of the QWs comprising the active zone [Delorme-2017b]. As observed, in general, there are no visible differences in the morphology of the QWs in the laser compared to the reference structure (cf. Fig. 19), which demonstrates that in spite of the growth challenges the degree of reproducibility is remarkable, further encouraging the use of Ga(Sb,Bi) for future optoelectronics devices.

Evaluation of the Ga(Sb,Bi)-on-GaSb interface width in the laser structure is shown in Fig. 20c and yields a chemical width of about 2.2 nm (defined by 10%–90% criterion) at the specific position of the line scan. As in previous cases, the interfaces are strikingly well defined by a sigmoidal function describing the intrinsic broadening at semiconductor heterointerfaces [Luna-2012] as observed in Fig. 20c which displays the experimental profile and the fitting to the sigmoidal function.

In analogy with the reference sample and in spite of the significant improvement with respect to the QWs in Fig. 16 and 17, slight fluctuations in QW thickness and interface width are still perceived in the laser structure. In this case, the interface width ranges between 2 and 2.8 nm, in the 10–90% criterion, without any signature of Bi surface segregation. Although the QWs are homogeneous in composition, the presence of lateral thickness fluctuations and, in particular, the non-steady interface width may impact the optical properties. This valuable information discloses the critical features to further improve the homogeneity of the QWs in terms of thickness and interface width. In any case, both Ga(Sb,Bi)-on-GaSb and GaSb-on-Ga(Sb,Bi) interfaces are rather symmetric and exhibit a very similar interface width.

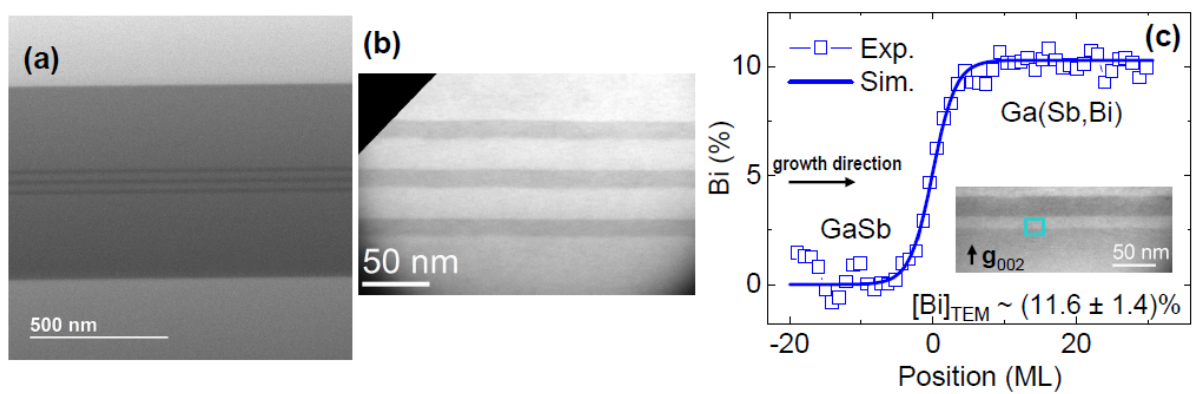

**Fig. 20** (a) Bright-field STEM overview of the laser structure with (b) an enlarged bright-field micrograph of the QWs comprising the active zone. (c) Bi distribution profile at the Ga(Sb,Bi)-on-GaSb interface at QWs in the laser structure, extracted from the analysis of the $\mathbf{g}_{002}$ DFTEM diffracted intensity. The experimental data are fitted to a sigmoidal function. Partly reprinted with permission from the American Institute of Physics [Luna-2018a], changed layout.

Finally, the abrupt interfaces in the laser structure also demonstrate that in-situ annealing during MBE growth of the top cladding and contact layers at the higher $T_s$ = 450 °C does not have a detrimental effect on the layers, at least in terms of interface quality.

# 4. Electronic band structure and optical properties

The electronic band structure of GaSbBi/GaSb QWs of various widths and contents was studied by photoreflectance (PR) [Kudrawiec-2018]. This method due to its absorption-like character probes energies of optical transitions between both the ground and excited states [Misiewicz-2012, Kudrawiec-2014]. From the comparison of PR data with calculations of energies of QW transitions performed for the varying valence band offset (VBO), it is possible to conclude about the number of confinement states and the VBO [Kudrawiec-2014]. In addition, PR is not sensitive to localized states in contrast to photoluminescence (PL) which probes the lowest energy states [Kopaczek-2015]. Therefore the optical quality of QWs can be evaluated by the comparison of PR and PL spectra and conclusions on carrier localization can be easily extracted from such a comparison.

## 4.1. Photoreflectance of GaSbBi/GaSb QWs

PR spectra measured at 10 K for nominally 11 and 15 nm wide $GaSb_{0.89}Bi_{0.11}$/GaSb QWs are shown in Fig. 21 a and b, respectively. QW widths and contents determined on the basis of TEM and XRD studies are given in the figure caption. For the two samples the strongest PR signal is observed at an energy of ~0.80 eV. This signal is associated with photon absorption in GaSb layers. The shape of this signal is different for the two samples due to various contributions from particular GaSb layers (cap, QW barriers, and buffer), which is typical for such structures. The PR features observed at energies lower than the GaSb signal are associated with the optical transitions in $GaSb_{0.89}Bi_{0.11}$/GaSb QWs. The fundamental transition is easy to identify since it is the PR resonance with the lowest energy and this resonance correlates very well with the PL peak which is shown by thick red line in Fig. 21. Due to the compressive strain present in these QWs the fundamental transition is between the heavy-hole subband and conduction subband. PR resonances, which are observed in the spectral range between the QW fundamental transition and the GaSb signal, are associated with the QW transitions between excited states. It is worth underlining here that the observation of optical transitions related to excited states is a clear experimental evidence that the studied QWs are of type I with a deep quantum confinement in the conduction and the valence band.

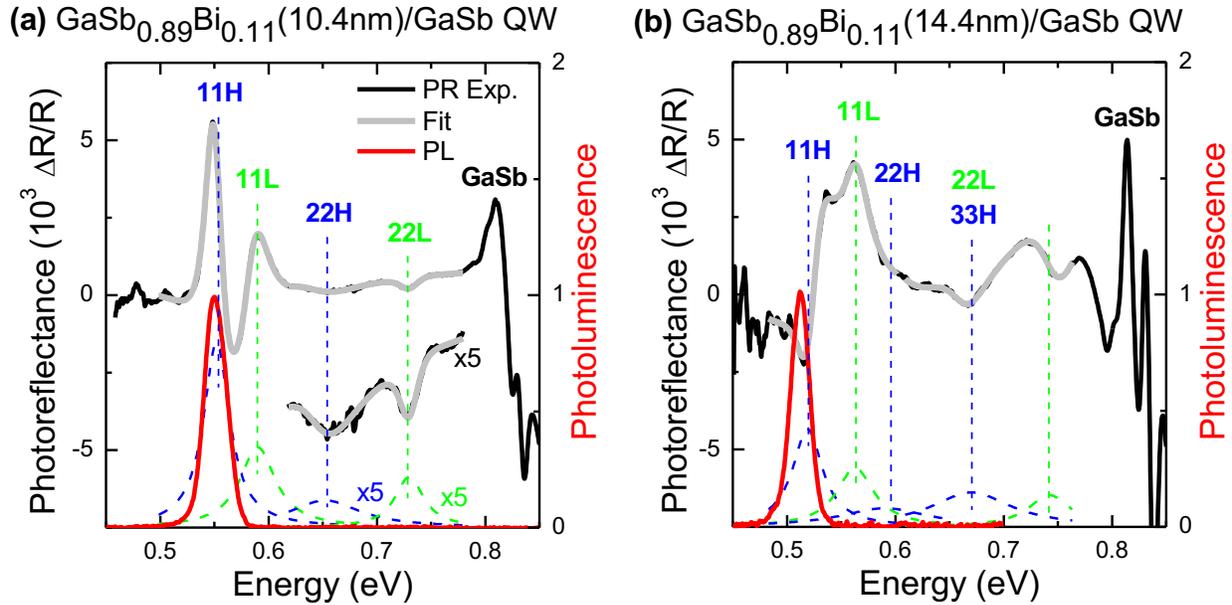

**Fig. 21** Photoreflectance (black lines) and photoluminescence (red lines) spectra of 10.4 (a) and 14.4 nm (b) wide GaSb$_{0.89}$Bi$_{0.11}$/GaSb QWs measured at 10 K. Thick solid grey lines represent theoretical fits and dashed lines correspond to the moduli of individual PR resonances (blue – heavy hole transitions, green – light hole transitions). From [Kudrawiec-2018].

The low-field electromodulation Lorentzian lineshape functional form [Aspnes-1973] was used to fit PR resonances and extract energies of QW transitions. The formula for the fitting is given below

$$\frac{\Delta R}{R}(E) = Re\left[\sum_{j=1}^{n} C_j\, e^{i\theta_j}(E - E_j + i\Gamma_j)^{-m_j}\right] \quad (3)$$

where $\frac{\Delta R}{R}(E)$ is the energy dependence of the PR signal, $n$ is the number of spectral functions to be fitted, $E$ is the photon energy of the probe beam, $E_j$ is the energy of the optical transition, and $\Gamma_j$, $C_j$ and $\theta_j$ are the broadening, amplitude and phase angle, respectively. The term $m_j$, which refers to the type of optical transitions, is assumed to be 2 since excitonic transitions are expected in this case at 10 K. The fitting curves are shown by grey lines in Fig. 21. The modulus of the individual PR resonance are obtained according to the following formula

$$\Delta\rho_j(E) = \frac{|C_j|}{\left[(E-E_j)^2 + \Gamma_j^2\right]^{\frac{m_j}{2}}} \quad (4)$$

with parameters taken from the fit. The modulus are shown by dashed lines in Fig. 21.

The identification of PR resonances was possible via a series of calculations [Kudrawiec-2018] and proper plots as shown in Fig. 22 a and b. The notation *nm*H(L) in Figs. 21 and 22 denotes the transition between *n*-th heavy-hole (light-hole) valence subband and *m*-th conduction subband.

According to the analysis shown in Fig. 22 the best correlation between the PR data and the theoretical calculations is observed for the VBO ~ 45-50% and PR resonances resolved in Fig. 21 are interpreted as follows. The resonance at the lowest energy is attributed to the 11H transition, which is the fundamental transition for the two $GaSb_{0.89}Bi_{0.11}$/GaSb QW samples. Moreover a 11L transition (*i.e.*, the fundamental transition for light-holes) is identified in the studied samples. In addition to the fundamental transition (11H and 11L transition), a 22H (22L) transition (*i.e.*, transition between the second heavy-hole (light-hole) subband and the second electron subband) is identified in PR spectra. For the sample with the 15 nm wide $GaSb_{0.89}Bi_{0.11}$/GaSb QW even the 33H transition is observed in PR spectrum. In this case the 33H transition overlaps with the 22L transition, but this is not a problem for the VBO analysis in this sample since the other transitions are well separated.

It is worth noting that the VBO determined in Fig. 22 is defined as

$$VBO = \frac{\Delta E_V}{(\Delta E_V + \Delta E_C)} \times 100\%, \quad (5)$$

where $\Delta E_V$ and $\Delta E_C$ are the valence- and conduction-band discontinuities at the heterojunction of unstrained materials. Since from a laser device perspective, the most interesting values are the band gap discontinuities with the strain corrections, the quantum confinement potential together with the electron and hole levels have been plotted in Fig. 22 c and d. In this case it is clearly visible that the quantum confinement potential for electrons and holes in the strained $GaSb_{0.89}Bi_{0.11}$/GaSb QWs is deep enough for laser applications.

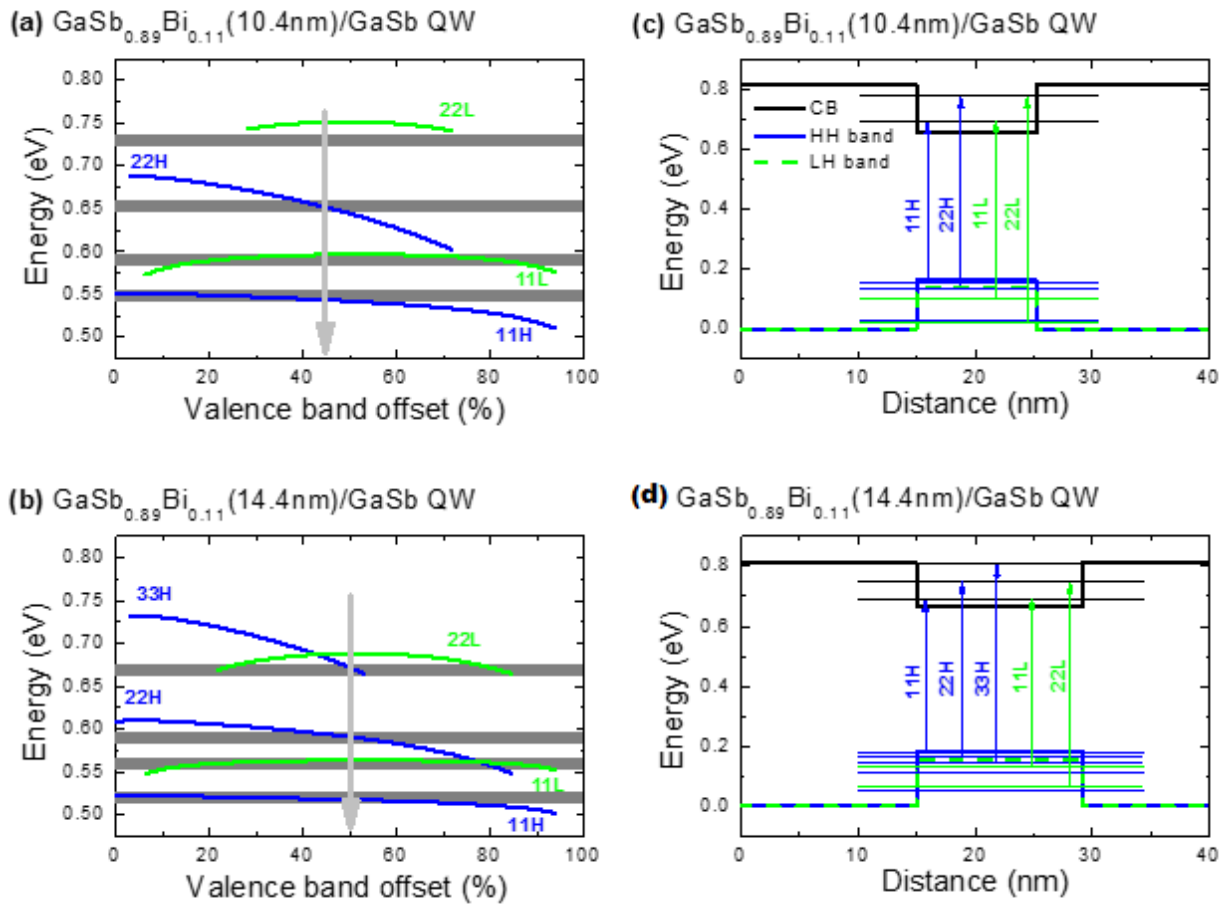

**Fig. 22** Method used to analyze the VBO in GaSb1-xBix/GaSb QWs: examples for 11 (a) and 15 nm (b) wide QWs with 11% Bi. The horizontal thick grey lines correspond to energies of 11H, 22H, 33H, 11L, and 22L transitions obtained from PR measurements and the solid lines represent theoretical calculations of energies of QW transitions for various values of VBO. Panels (c) and (d) show the quantum confinement potential for the QWs analyzed in panels (a) and (b) together with energy levels. From [Kudrawiec-2018].

The VBO extracted from the comparison of PR data with theoretical calculations is very consistent with the VBO determined from *ab initio* calculations [Polak-2017, Kudrawiec-2018]. These calculations show that the incorporation of Bi atoms into a GaSb host modifies both the conduction band (CB) and valence band (VB). The variation rates are very similar for CB (~15-16 meV per % Bi) and for VB (~15-16 meV per % Bi), which, in consequence, leads to a reduction rate of the band gap of ~30-32 meV per % Bi. It gives the ~48-52 % VBO between $GaSb_{1-x}Bi_x$ and GaSb.

In Ref. [Kudrawiec-2018], it has been also concluded that the electron effective mass reduces linearly with the increase in Bi concentration (x):

$$m_{eff}^{GaSbBi} = m_{eff}^{GaSb} - 0.2x, \tag{6}$$

where $m_{eff}^{GaSb}$ is the electron effective mass in GaSb. Moreover the effective mass of holes changes with Bi concentration but this effect is weaker than the Bi-related change in the electron effective mass and therefore it can be neglected at the first approximation.

Photoreflectance has been also applied to study the electronic band structure of GaSb(Bi)/AlGaSb QWs. In this case the authors have identified optical transitions related to both the ground and excited states [Chen-2015]. They have been observed that these transitions shift to red due to incorporation of Bi atoms into the QW region.

## 4.2. Photoluminescence of GaSbBi/GaSb QWs and carrier localization

A strong carrier localization is generally observed with dilute bismides [Kopaczek-2015, Kudrawiec-2009, Shakfa-2013, Fitouri-2015, Gelczuk-2017]. This phenomenon is typical of HMA and is associated with alloy inhomogeneities and other imperfections like point defects etc. [Gelczuk-2017]. In general, carrier localization occurs at low temperatures and can be treated as an indicator of sample (material) quality. Strong photoluminescence from 6-7nm wide GaSbBi/GaSb QWs with varying Bi concentration was reported by Zhang et al. [Zhang-2018]. For these samples, carrier localization was weak at low temperatures, quite strong photoluminescence was observed up to the room temperature, and the room temperature emission was associated to recombination between delocalized states. Strong room temperature photoluminescence was also reported for delta doped GaSbBi QWs by Yue at al. [Yue-2018b]. These results suggest that good quality QWs can be achieved with this alloy.

For the samples shown in Fig. 21 the Stokes shift (*i.e.*, the energy difference between absorption and emission) is negligible at 10 K under the given excitation density, which in this case equals ~100 W/cm$^2$. For the two QW samples analyzed in Fig. 21, the energy position of the PL peak does not change when changing the excitation power by four magnitudes (Fig. 23a). Moreover the temperature dependence of the PL peak position does not show the S-shape behavior typical of carrier localization (Fig. 23b). This means that carrier localization in GaSbBi/GaSb QWs grown under optimal conditions is weak or even negligible.

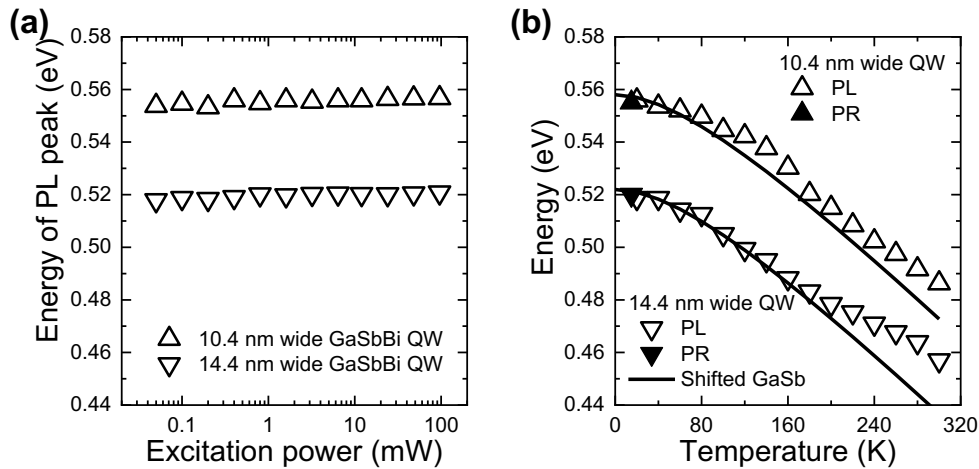

**Fig. 23** PL peak position for 10.4 (up triangles) and 14.4 nm (down triangles) wide $GaSb_{0.89}Bi_{0.11}$/GaSb QWs for various excitation powers at 10K (a) and various temperatures at the excitation power of 100 mW (b). Solid lines correspond to the temperature dependence of GaSb bandgap described by Varshni formula with α = 0.417 mV/K and β =140 K [Vurgaftman-2001] and shifted the energy of fundamental transition determined from PR measurements. Solid triangles correspond to the energy of the ground state transition determined by PR measurements. From [Linhart-2018].

In general, strong carrier localization is an intrinsic property of HMAs including dilute bismides [Kopaczek-2015, Kudrawiec-2009, Shakfa-2013, Fitouri-2015, Gelczuk-2017], and it is difficult or almost impossible to eliminate this phenomenon. However in the GaSbBi case one deals with mixing similar atoms (Sb and Bi) in terms of their electro-negativities and sizes, in contrast to GaAsBi. Therefore the optical quality of GaSbBi/GaSb QWs can be much better than GaAsBi/GaAs QWs, and GaSbBi alloy can be treated as a regular III-V alloy (*i.e.*, like GaInSb or GaPAs). This conclusion is derived from PL studies of $GaSb_{1-x}Bi_x$/GaSb QWs [Linhart-2018] and is very consistent with theoretical predictions obtained on the basis of the analysis of band anticrossing parameters which describe the highly mismatched alloys [Polak-2015]. Therefore we believe that good quality GaSbBi/GaSb QWs can be grown even with Bi contents and QW width close to the critical thickness, *i.e.* ~25% Bi for 11-15 nm wide QWs.

# 5. Laser based on GaSbBi/GaSb quantum-wells

## 5.1. Laser structure

The demonstration of a laser structure based on GaSbBi quantum wells (QWs) was recently realized [Delorme-2017b], and is described in the following. The laser structure is shown in Fig. 24a. The active region was sandwiched between 263 nm-thick GaSb waveguide layers and 1.65 µm thick $Al_{0.8}Ga_{0.2}AsSb$ cladding layers, and a 300-nm-thick highly p-type doped GaSb:Be was used as a top contact layer. The GaSbBi/GaSb QW active region is composed of three 15 nm-GaSbBi/20 nm-GaSb type-I QWs. Fig. 24b shows the band alignment and the energy position of electrons and holes levels for the GaSbBi/GaSb QW calculated using the nextnano© suite [NEXTNANO]. Bi-related changes in the conduction band and valence band positions of GaSbBi were taken from recent theoretical studies using the valence-band anticrossing (VBAC) and virtual-crystal approximation (VCA) models [Polak-2015, Samajdar-2016]. The energy separation at room temperature of the first electron- and hole- confined levels is about 0.45 eV, and both holes and electrons are well confined in the GaSbBi QW.

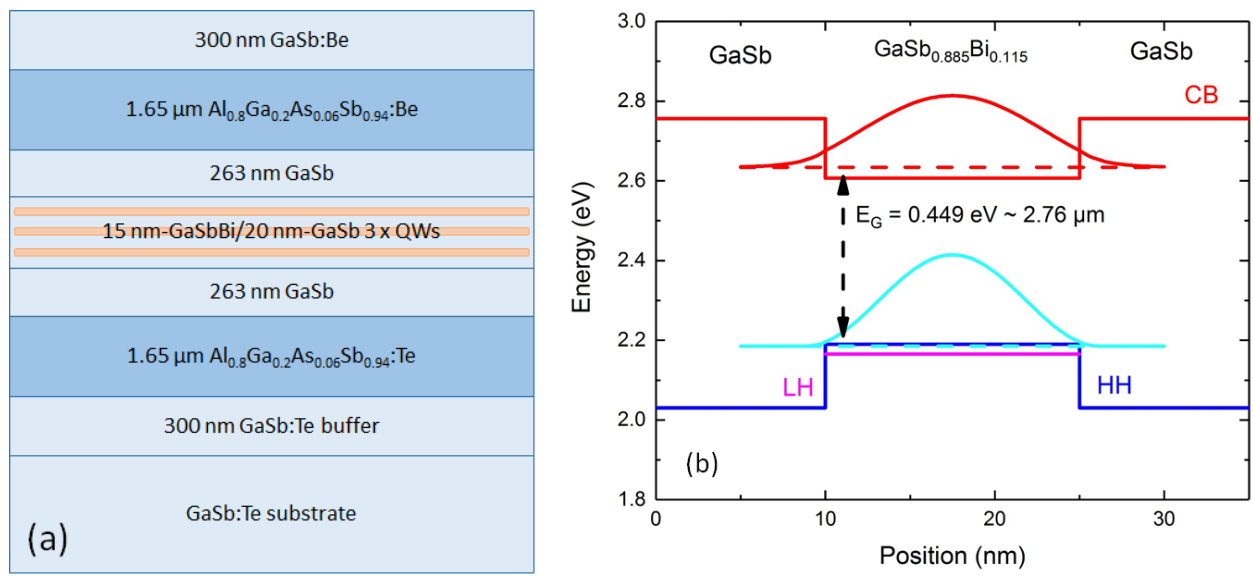

**Fig. 24** (a) Schematic view of the GaSbBi/GaSb MQW laser structure. (b) Calculated band alignment of 15-nm $GaSb_{0.885}Bi_{0.115}$/GaSb QW. Reprinted with permission from the American Institute of Physics [Delorme-2017b], changed layout.

## 5.2. Epitaxial growth of the laser structure

The V/III flux ratio was set by growing a 110-nm thick GaSbBi epilayer at 200°C (substrate heater thermocouple reading (TTR)). Under the same conditions, a GaSbBi/GaSb QW structure was grown prior to the laser diode, using a similar active region. Due to the very low growth temperature, a V/III flux ratio close to stoichiometry was also used during GaSb barrier-layer growth to avoid deposition of metallic Sb on the surface. The QW were embedded between 180 nm thick GaSb layers and 20 nm thick AlAs$_{0.08}$Sb$_{0.92}$ barrier layers lattice-matched to the substrate to confine the optically generated carriers during PL spectroscopy. These GaSb and AlAsSb layers were grown at 425°C (pyrometer reading). The structure was completed by a 20 nm thick GaSb cap layer to avoid oxidation of the topmost AlAsSb layer.

Finally, the laser diode was grown. The active region was grown at 200°C (TTR), whereas the other layers were grown at 450°C. Fig. 25 shows the HR-XRD $\omega$–$2\theta$ scans measured on the three samples. Simulations indicate a Bi concentration as high as 12.8% in the 110 nm thick layer. In addition, the excellent crystal quality is demonstrated by well-defined Pendellösung fringes and the perfect agreement between simulated and experimental curves. Similar conclusions – high crystal quality and sharp interfaces – can be drawn for the GaSbBi/GaSb MQW structure. The laser structure exhibits slightly broader features near the substrate peak, probably due to a slight mismatch of the cladding layers. Nonetheless, the overall experimental curve is in excellent agreement with the simulation. Notably, the GaSbBi/GaSb QW-related features are clearly visible which shows that the whole periodicity has been preserved. An interesting point is that HR-XRD simulations indicate that the Bi-content in both the MQW and the laser structures is 11.5% for both samples, i.e. slightly lower than in the 110 nm single layer grown under the same conditions. This could arise from a lower incorporation rate of Bi in the first couple of nanometers of GaSbBi growth, as previously reported in the case of GaAsBi [Fan-2013, Makhloufi-2014]: it was indeed proposed that the Bi acts as a surfactant during the growth of the first nanometers, until a Bi-rich surface is observed, which slightly lowers the Bi content at the very beginning of the layer.

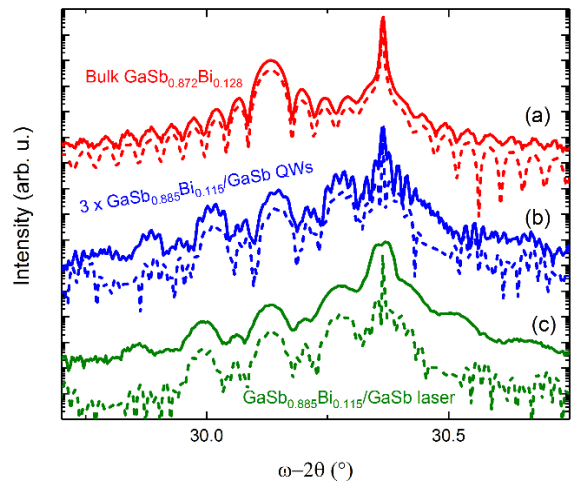

**Fig. 25** HR-XRD (004) ω–2θ scans of (a) the GaSb$_{0.872}$Bi$_{0.128}$ single layer (solid red line), (b) the GaSb$_{0.885}$Bi$_{0.115}$/GaSb MQW structure (solid blue line), and (c) the laser structure (solid green line). The simulated curves are given as dashed lines. The curves have been vertically shifted for clarity. Reprinted with permission from the American Institute of Physics [Delorme-2017b].

### 5.3. Device characterization

Ridge laser diodes of 10x1160 µm² were processed using standard photolithography and wet etching. Ti-Au and AuGeNi were used as contact metals for the p- and n-type contacts, respectively. Electrical insulation and protection of the etched sidewalls was obtained using the AZ1518 photoresist. Laser cavities were formed by simple cleaving of the facets, without any antireflection coating. Finally, the devices were soldered epi-side down with indium on Cu heat-sinks. Next, the laser was characterized at different temperatures under pulsed injection (200 ns pulse width, 21 kHz repetition rate). Fig. 26 shows the light-current (L-I) and voltage-current (V-I) characteristics (a) and the laser emission spectrum at different temperatures measured for an injected current slightly above threshold (b). The V-I characteristic measured at RT clearly shows a diode behavior with a turn-on voltage close to 0.7 V and a threshold voltage of 1.25 V. At RT, the threshold current density $J_{th}$ for the 10x1160 µm² area diode is 4.22 kA/cm² with a lasing wavelength of 2.71 µm (0.457 eV) under pulsed operation. At 80 K, the threshold current density is 431 A/cm² and the emitted wavelength 2.50 µm. The $T_0$ characteristic temperature of this laser diode is 111 K between 80 and 250 K and decreases to 53 K in the 250 to 300 K temperature range.

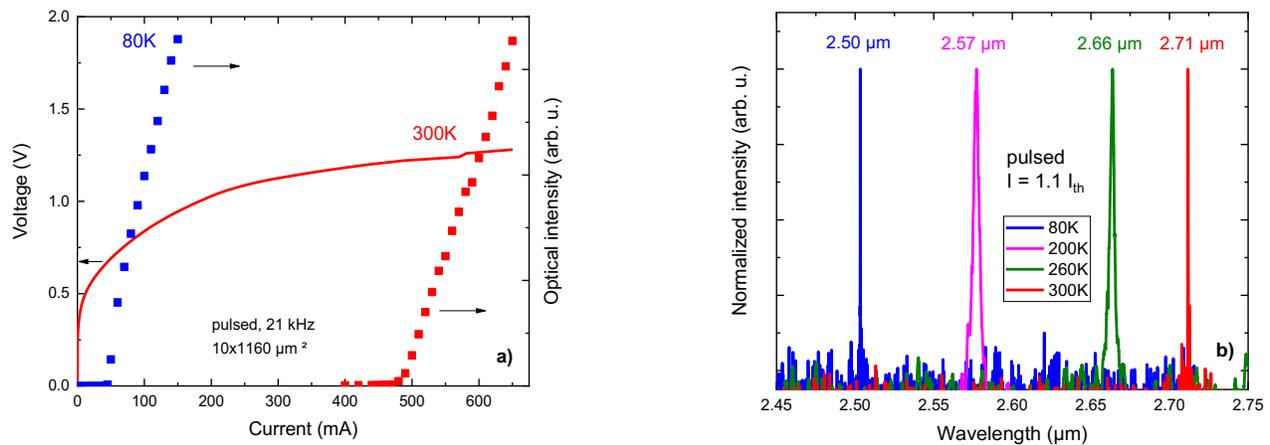

**Fig. 26** (a) I-V and L-I (under pulsed operation) characteristics at RT (in red) and 80K (in blue). (b) Laser emission spectrum at different temperatures under pulsed operation. The intensities are normalized to 1. Reprinted with permission from the American Institute of Physics [Delorme-2017b].

Finally, CW operation was achieved at 80 K from this laser structure with a lasing wavelength of 2.52 µm (Fig. 27) and a threshold current density $J_{th}$ of 586 A/cm². The $J_{th}$ of the device is still relatively high compared to the standard GaInAsSb/AlGaAsSb laser diodes developed for many years in this spectral range but is comparable to those of early laser diodes [Choi-1994, Lee-1995, Garbuzov-1995]. This high threshold current density may be related to a large density of radiative defects, which would be consistent with the relatively broad PL peaks observed, and investigation are thus required to further improve the GaSbBi / GaSb QW structural quality.

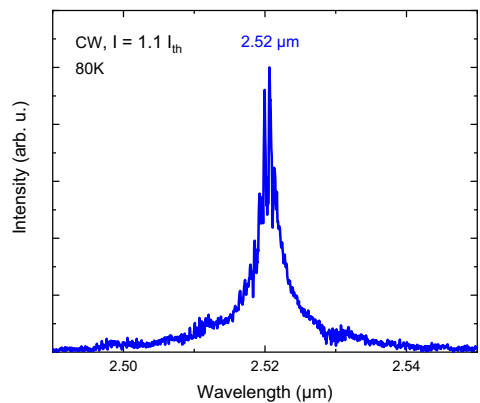

**Fig. 27** CW laser emission at 80K. Reprinted with permission from the American Institute of Physics [Delorme-2017b].

## 6. Conclusion

Recent advances in the fabrication and characterization of GaSbBi alloys have been reviewed in this chapter. The potential use of this alloy for new mid-IR device designs has fostered the development and a better understanding of the epitaxial growth, which resulted in the realization of homogeneous, high-quality bulk layers and quantum-wells. The microstructure analysis and optical studies carried out recently on this material and supporting this statement were described in this chapter. Despite the challenging growth conditions (very low substrate temperature, V/III flux ratio close to unity), a complete laser structure was fabricated and actual devices could be processed. The material quality as well as the heterostructure design are certainly still to be improved in order to bring lasers performance to state-of-the-art. Nevertheless, this first demonstration, together with the recent material assessments confirm that GaSbBi is a promising material for future mid-IR device development.

## Acknowledgements

The work at the University of Montpellier was partly supported by the French program "Investments for the Future" (EXTRA, ANR-11-EQPX-0016) and the national research agency (BIOMAN, ANR-15-CE24-0001).

# References

[Abedin-2006] M. N. Abedin *et al.* "Progress of Multicolor Single Detector to Detector Array Development For Remote Sensing", Proc. SPIE 5543, 239 (2004).

[Ashwin-2011] M. J. Ashwin *et al.* "Controlled nitrogen incorporation in GaNSb alloys" AIP Adv. 1, 032159 (2011).

[Aspnes-1973] D. E. Aspnes "Third-derivative modulation spectroscopy with low-field electroreflectance" Surf. Sci. 37, 418 (1973).

[Bahriz-2015] M. Bahriz *et al.* "High temperature operation of far infrared (λ~20µm) InAs/AlSb quantum cascade lasers with dielectric waveguide" Opt. Exp., vol. 23 (2), pp. 1523-1528, 2015

[Balades-2018] N. Baladés *et al.* "Analysis of Bi Distribution in Epitaxial GaAsBi by Aberration-Corrected HAADF-STEM" Nanoscale Research Letters 13:125 (2018).

[Bastiman-2012] F. Bastiman *et al.* "Bi incorporation in GaAs(100)-2×1 and 4×3 reconstructions investigated by RHEED and STM" J. Cryst. Growth 341 19, (2012).

[Beyer-2017] A. Beyer *et al.* "Local Bi ordering in MOVPE grown Ga(As,Bi) investigated by high resolution scanning transmission electron microscopy" Applied Materials Today **6**, 22–28 (2017).

[BIANCHO-2010] BIsmide And Nitride Components for High temperature Operation – European Project FP7-STREP n°257974 -07/2010 – 06/2013 – www.biancho.org

[Bithell-1989] E.G. Bithell and W.M. Stobbs "Composition determination in the GaAs/(Al, Ga) As system using contrast in dark-field transmission electron microscope images" Phil. Mag. A 60 39, (1989)

[Butkute-2012] R. Butkutė et al. "Thermal annealing effect on the properties of GaBiAs" Phys. Status Solidi C 9, 1614 (2012).

[Butkute-2017] R. Butkutė *et al.* "Bismuth quantum dots in annealed GaAsBi/AlAs quantum wells" Nanoscale Research Letters 12:436, 7 pp, (2017).

[Castellano-2017] A. Castellano *et al.* "Room-temperature continuous-wave operation in the telecom wavelength range of GaSb-based lasers monolithically grown on Si" APL Photonics 2, 061301 (2017).

[Cerutti-2015] L. Cerutti *et al.* "GaSb-based composite quantum wells for laser diodes operating in the telecom wavelength near 1.55µm" Appl. Phys. Lett., vol. 106, pp. 101102, 2015

[Chen-2008] J. Chen *et al.*, "Effect of quantum well compressive strain above 1% on differential gain threshold current density in type-I GaSb-based diode lasers", IEEE J. of Quant. Electr., vol. 44 (12), pp. 1204-1201, 2008 [Chen-2015] X.R. Chen *et al.* "Bismuth Effects on Electronic Levels in GaSb(Bi)/AlGaSb Quantum Wells Probed by Infrared Photoreflectance" Chinese Physics Letters 32, 067301 (2015).

[Choi-1994] H. K. Choi *et al.* "Double-heterostructure diode lasers emitting at 3 µm with a metastable GaInAsSb active layer and AlGaAsSb cladding layers" Appl. Phys. Lett. 64, 2474 (1994).

[Das-2012] S.K. Das *et al.* "Near infrared photoluminescence observed in dilute GaSbBi alloys grown by liquid phase epitaxy", Infrared Phys. & technol., vol. 55 (1), pp. 156-160, 2012

[Delorme-2017a] O. Delorme *et al.* "Molecular beam epitaxy and characterization of high Bi content GaSbBi alloys" J. Cryst. Growth, 477 (2017), pp. 144-148

[Delorme-2017b] O. Delorme *et al.* "GaSbBi/GaSb quantum well laser diodes" Appl. Phys. Lett. 110 (2017) 222106.

[Delorme-2018] O. Delorme *et al.* "In situ determination of the growth conditions of GaSbBi alloys" J. Cryst. Growth 2018, 495, 9–13.

[Doyle-1968] P.A Doyle and P. S. Turner "Relativistic Hartree-Fock X-ray and electron scattering factors" Acta Crystallogr., Sect. A 24, 390 (1968)

[Duzik-2012] A. Duzik et al. "Surface structure of bismuth terminated GaAs surfaces grown with molecular beam epitaxy" Surf. Sci. 606, 1203 (2012).

[Duzik-2014] A. Duzik *et al.* "Surface morphology and Bi incorporation in GaSbBi(As)/GaSb films" " J. Cryst. Growth, vol. 390, pp. 5-11, 2014

[Fan-2013] D. Fan *et al.* "MBE grown GaAsBi/GaAs double quantum well separate confinement heterostructures" J. Vac. Sci. Technol. B: Microelectron. Nanometer Struct., 31 (3) (2013), pp. 181103-181106

[Ferhat-2006] M. Ferhat, A. Zaoui, "Structural and electronic properties of III-V bismuth compounds", Phys. Rev. B 73, 115107 (2006).

[Fitouri-2015] H. Fitouri *et al.* "Photoreflectance and photoluminescence study of localization effects in GaAsBi alloys" Optical Materials 42, 67 (2015).

[Fuyuki-2014] T. Fuyuki *et al.* "Electrically pumped room-temperature operation of GaAs$_{1-x}$Bi$_x$ laser diodes with low-temperature dependence of oscillation wavelength" Appl. Phys. Express, vol. 7, pp. 082101, 2014

[Garbuzov-1996] D. Z. Garbuzov *et al.* "2.7-µm InGaAsSb/AlGaAsSb laser diodes with continuous-wave operation up to −39 °C" Appl. Phys. Lett. 67, 1346 (1995).

[Gardes-2014] C. Gardes *et al.* "100nm AlSb/InAs HEMT for ultra-low power consumption, low-noise applications" Scientific world Journal, Art. Num. 136340, 2014 [Gelczuk-2017] L. Gelczuk *et al.* "Deep-level defects in n-type GaAsBi alloys grown by molecular beam epitaxy at low temperature and their influence on optical properties" Scientific Reports 7, 2824 (2017).

[Glas-2004] F. Glas et al. "Determination of the Local Concentrations of Mn Interstitials and Antisite Defects in GaMnAs" Phys. Rev. Lett. 93, 086107 (2004).

[Godoy-2015] S. E. Godoy *et al.* "Dynamic infrared imaging for skin cancer screening" Infrared Physics & Technology. 2015;70:147–152

[Gu-2014] Y. Gu *et al.*, "Structural and optical characterizations of InPBi thin films grown by molecular beam epitaxy" Nanoscale. Res. Lett., vol. 9, 2014

[Gurjarpadhye-2015] A. A. Gurjarpadhye *et al.* "Infrared Imaging Tools for Diagnostic Applications in Dermatology" SM J Clin Med Imaging. 2015 ; 1(1): 1–5.

[Hosoda-2010] T. Hosoda *et al.* "Type-I GaSb-based laser diodes operating in 3.1 to 3.3 µm wavelength range", IEEE Phot. Tech. Lett., vol. 22 (10), pp. 718-720, 2010

[Jacobsen-2012] H. Jacobsen et al. "*Ab initio* study of the strain dependent thermodynamics of Bi doping in GaAs" Phys. Rev. B 86, 085207 (2012).

[Janotti-2002] A. Janotti *et al.*, "Theoretical study of the effects of isovalent coalloying of Bi and N in GaAs" Phys. Rev. B 65, 115203 (2002)

[Joukoff-1972] B. Joukoff *et al.* "Growth of $InSb_{1-x}Bi_x$ single crystals by Czochralski method", J. Cryst. Growth, vol. 12 (2), pp. 169-172, 1972

[Kopaczek-2014] J. Kopaczek *et al.* "Photoreflectance spectroscopy of GaInSbBi and AlGaSbBi quaternary alloys" Appl. Phys. Lett., vol. 105 (11), pp. 112102, 2014

[Kopaczek-2015] J. Kopaczek *et al.* "Optical properties of GaAsBi/GaAs quantum wells: Photoreflectance, photoluminescence and time-resolved photoluminescence study", Semicond. Sci. Technol. 30, 094005 (2015).

[Kudrawiec-2009] R. Kudrawiec *et al.* "Carrier localization in GaBiAs probed by photomodulated transmittance and photoluminescence" J. Appl. Phys. 106, 023518 (2009).

[Kudrawiec-2014] R. Kudrawiec *et al.* "Experimental and theoretical studies of band gap alignment in $GaAs_{1-x}Bi_x$/GaAs quantum wells", J. Appl. Phys. 116, 233508 (2014).

[Kudrawiec-2018] R. Kudrawiec et al., "Type I $GaSb_{1-x}Bi_x$/GaSb quantum wells dedicated for mid infrared laser applications: Photoreflectance studies of band gap alignment", Phys. Rev. Appl. submitted (2018).

[Lafaille-2012] P. Lafaille *et al.* "High temperature operation of short wavelength InAs-based quantum cascade lasers", AIP Advances, vol. 2 (2), pp. 02219, 2012

[Lee-1995] H. Lee *et al.* "Room-temperature 2.78 μm AlGaAsSb/InGaAsSb quantum-well lasers" Appl. Phys. Lett. 66, 1942 (1995).

[Lewis-2012] R.B. Lewis *et al.* "Growth of high Bi concentration GaAs$_{1-x}$Bi$_x$ by molecular beam epitaxy" Appl. Phys. Lett., vol. 100 (5), pp. 082112, 2012

[Linhart-2017] W. Linhart *et al.* "Indium-incorporation enhancement of photoluminescence properties of Ga(In)SbBi alloys" J. Phys. D: Appl. Phys. 50 (2017) 375102

[Linhart-2018] W. Linhart *et al.* "Weak carrier localization in GaSbBi/GaSb QWs studied by photoluminescence and time resolved photoluminescence" Appl. Phys. Lett. to be submitted (2018).

[Liu-2008] C. Liu *et al.*, "Quantum Spin Hall Effect in Inverted Type-II Semiconductors", Phys. Rev. Lett. 100, 236601 (2008)

[Lu-2015] J. Lu *et al.* "Investigation of MBE-grown InAs$_{1-x}$Bi$_x$ alloys and Bi-mediated type-II superlattices by transmission electron microscopy" J. Cryst. Growth 425, 250 (2015).

[Lu-2016] J. Lu et al. "Evaluation of antimony segregation in InAs/InAs$_{1-x}$Sb$_x$ type-II superlattices grown by molecular beam epitaxy" J. Appl. Phys. 119, 095702 (2016).

[Ludewig-2013] P. Ludewig *et al.* "Electrical injection Ga(AsBi)/(AlGa)As single quantum well laser" Appl. Phys. Lett., vol. 102 (24), pp. 242115, 2013

[Luna-2008] E. Luna et al. "Indium distribution at the interfaces of (Ga,In)(N,As)/GaAs quantum wells" Appl. Phys. Lett. 92, 141913 (2008).

[Luna-2009] E. Luna et al. "Interface properties of (Ga,In)(N,As) and (Ga,In)(As,Sb) materials systems grown by molecular beam epitaxy" J. Cryst. Growth 311, 1739 (2009).

[Luna-2012] E. Luna, A. Guzmán et al. "Critical role of two-dimensional island-mediated growth on the formation of semiconductor heterointerfaces" Phys. Rev. Lett. 109, 126101 (2012).

[Luna-2015] E. Luna et al. "Spontaneous formation of nanostructures by surface spinodal decomposition in GaAs$_{1-x}$Bi$_x$ epilayers" J. Appl. Phys. 117 185302, (2015).

[Luna-2016] E. Luna et al. "Spontaneous formation of three-dimensionally ordered Bi-rich nanostructures within GaAs$_{1-x}$Bi$_x$/GaAs quantum wells" Nanotechnology 27 (32), 325603, (2016)

[Luna-2017] E. Luna et al. "Morphological and chemical instabilities of nitrogen delta-doped GaAs/(Al, Ga) As quantum wells" Appl. Phys. Lett. 110, 201906 (2017).

[Luna-2018a] E. Luna et al. "Microstructure and interface analysis of emerging Ga (Sb, Bi) epilayers and Ga (Sb, Bi)/GaSb quantum wells for optoelectronic applications" Appl. Phys. Lett. 112, 151905 (2018)

[Luna-2018b] E. Luna et al. "Transmission Electron Microscopy of Ga(Sb,Bi)/GaSb Quantum Wells with varying Bi content and quantum well thickness" submitted Semicond. Sci. Technol. (2018)

[Makhloufi-2014] H. Makhloufi *et al.* "Molecular beam epitaxy and properties of GaAsBi/GaAs quantum wells grown by molecular beam epitaxy: effect of thermal annealing" Nanoscale. Res. Lett., vol. 9, 2014

[Marko-2015] I. P. Marko *et al.*, "Properties of hybrid MOVPE/MBE grown GaAsBi/GaAs based near-infrared emitting quantum well lasers" Semicond. Sci. Technol. 30, 094008–0910 (2015).

[Matthews-1974] J. W. Matthews and A. E. Blakeslee "Defects in epitaxial multilayers" J. Cryst. Growth 27, 118 (1974)

[Mazzucato-2013] S. Mazzucato *et al.* "reduction of defect density by rapid thermal annealing in GaAsBi studied by time-resolved photoluminescence" Semicond. Sci. Technol., vol. 28 (2), pp. 022001, 2013

[Misiewicz-2012] J. Misiewicz, R. Kudrawiec, "Contactless electroreflectance spectroscopy of optical transitions in low dimensional semiconductor structures", Opto-Electronics Review 20, 101 (2012).

[Mohmad-2012] A.R. Mohmad *et al.* "Effect of rapid thermal annealing on $GaAs_{1-x}Bi_x$ alloys", Appl. Phys. Lett., vol. 101 (1), pp. 012106, 2012

[Muraki-1992] K. Muraki et al. "Surface segregation of In atoms during molecular beam epitaxy and its influence on the energy levels in InGaAs/GaAs quantum wells" Appl. Phys. Lett. 61, 557 (1992)

[NEXTNANO] See www.nextnano.com/index.php for "Nextnano GmbH - semiconductor software solutions."

[Nguyen-Van-2018] Nguyen-Van et al. "Quantum cascade lasers grown on silicon" Scientific Reports 8, 7206 (2018).

[Noreika-1982] A. J. Noreika *et al.* "Indium antimonide-bismuth compositsions grown by molecular beam epitaxy", J. of Appl. Phys., vol. 53 (7), pp. 4932-4937, 1982

[Norman-2011] A.G. Norman et al. "Atomic ordering and phase separation in MBE $GaAs_{1-x}Bi_x$" J. Vac. Sci. Technol. B 29, 03C121, (2011)

[Oe-2002] K. Oe "Metalorganic vapour phase epitaxial growth of metastable $GaAs_{1-x}Bi_x$ alloy", J. Cryst. Growth, vol. 237, pp.1481-1485, 2002

[O'Malley-2010] R. O'Malley *et al.*, "Detection of pedestrians in far-infrared automotive night vision using region-growing and clothing distortion compensation", Infrared Physics and Technology 53 (2010) 439-449

[Pan-2000] Z. Pan et al. "Kinetic modeling of N incorporation in GaInNAs growth by plasma-assisted molecular-beam epitaxy", Appl. Phys. Lett. 77, 214 (2000).

[Patil-2017] P.K. Patil et al. "GaAsBi/GaAs multi-quantum well LED grown by molecular beam epitaxy using a two-substrate-temperature technique" Nanotechnology 28(10) 105702, (2017)

GaSbBi alloys and heterostructures: fabrication and properties                                                                 46[Pecharoman-2013] R. Pecharoman-Gallego, "Quantum cascade lasers: review, applications and prospective development", Lasers in engineering, vol. 24 (5-6), pp. 277-314, 2013

[Polak-2014] M.P. Polak *et al.* "Theoretical and experimental studies of electronic band structure for GaSb$_{1-x}$Bi$_x$ in the dilute Bi regime" J. of Phys. D: Appl. Phys., vol. 47 (35), pp. 355107, 2014

[Polak-2015] M. Polak *et al.* "First-principles calculations of bismuth induced changes in the band structure of dilute Ga–V–Bi and In–V–Bi alloys: chemical trends versus experimental data" Semicond. Sci. Technol. 30, 094001 (2015).

[Punkkinen-2015] M.P.J. Punkkinen et al. "Thermodynamics of the pseudobinary GaAs$_{1-x}$Bi$_x$ (0 ≤ x ≤ 1) alloys studied by different exchange-correlation functionals, special quasi-random structures and Monte Carlo simulations" Comput. Condens. Matter 5, 7, (2015)

[Puustinen-2013] J. Puustinen *et al.* "Variation of lattice constant and cluster formation in GaAsBi", J. of Appl. Phys., vol. 114 (24), pp. 243504, 2013

[Rajpalke-2013] M. K. Rajpalke *et al.* "Growth and properties of GaSbBi alloys" Appl. Phys. Lett. 103, 142106 (2013).

[Rajpalke-2014] M. K. Rajpalke *et al.* "High Bi content GaSbBi alloys" J. of Appl. Phys., vol. 116 (4), pp. 043511, 2014

[Rajpalke-2015] M. K. Rajpalke *et al.* "Bi flux-dependent MBE growth of GaSbBi alloys" J. Cryst. Growth 2015, 425, 241–244.

[Razhegi-2014] M. Razeghi *et al.* "Advances in mid-infrared detection and imaging: a key issues review" Rep. Prog. Phys. 77 (2014) 082401

[Reboul-2011] J.R. Reboul *et al.*, "Continuous wave operation above room temperature of GaSb-based laser diodes grown on Si", Appl. Phys. Lett., vol. 99 (12), pp. 121113, 2011

[Reyes-2014] D.F. Reyes *et al.*, "Bismuth incorporation and the role of ordering in GaAsBi/GaAs structures", Nanoscale. Res. Lett., vol. 9, 2014

[Rhiger-2011] D. R. Rhiger, "Performance Comparison of Long-Wavelength Infrared Type II Superlattice Devices with HgCdTe" J. Electron. Mater. 40, 1815 (2011).

[Rothman-09] L. S. Rothman *et al.*, "The HITRAN 2008 molecular spectroscopic database," J. Quant. Spectrosc. Radiat. Transf. 110(9-10), 533-572 (2009).

[Sales-2011] D. L. Sales et al. "Distribution of bismuth atoms in epitaxial GaAsBi" Appl. Phys. Lett. 98, 101902, (2011).

[Samajdar-2014] D. P. Samajdar *et al.*, "Calculation of direct E0 energy gaps for III-V-Bi alloys using quantum dielectric theory" In *The Book Physics of Semiconductor Devices*; Springer International Publishing: Cham, Switzerland, 2014; pp. 779–781.

GaSbBi alloys and heterostructures: fabrication and properties                                                      47[Samajdar-2016] D. P. Samajdar *et al.* "Calculation of Valence Band Structure of GaSb1–xBix Using Valence Band Anticrossing Model in the Dilute Bi Regime" *Recent Trends in Materials and Devices* (Springer International Publishing, 2016), pp. 243–248.

[Sandall-2014] I. Sandall *et al.* "Demonstration of InAsBi photoresponse beyond 3.5 µm" Appl. Phys. Lett., vol. 104 (17), pp. 171109, 2013

[Shakfa-2013] M. K. Shakfa *et al.* "Quantitative study of localization effects and recombination dynamics in GaAsBi/GaAs single quantum wells" J. Appl. Phys. 114, 164306 (2013).

[Song-2012] Y. Song *et al.* "Growth of GaSb1-xBix by molecular beam epitaxy" J. of Vac. Sc. and Technol. B, vol. 30 (2), pp. 02B114, 2012

[Steele-2016] J.A. Steele et al. "Surface effects of vapour-liquid-solid driven Bi surface droplets formed during molecular-beam-epitaxy of GaAsBi" Scientific Reports 6, 28860 (2016).

[Tait-2016] C.R. Tait and J.M. Millunchick "Kinetics of droplet formation and Bi incorporation in GaSbBi alloys" J. Appl. Phys. **119**, 215302, (2016).

[Tait-2017] C.R. Tait et al. "Droplet induced compositional inhomogeneities in GaAsBi" Appl. Phys. Lett. 111, 042105 (2017).

[Tan-2018] B.-S. Tan *et al.* "The 640×512 LWIR type-II superlattice detectors operating at 110 K" Infrared Phys. Techn. 89 (2018) 168-173

[Tidrow-2001] M.Z. Tidrow *et al.*, "Infrared sensors for ballistic missile defense", Infrared Physics and Technology, Vol 42, Issue 3-5, p.333 (2001)

[Tiedje-2008] T. Tiedje et al. "Growth and properties of the dilute bismide semiconductor GaAs$_{1-x}$Bi$_x$ a complementary alloy to the dilute nitrides" Int. J. Nanotechnol. **5**, 963 (2008).

[Tittel-2013] F.K. Tittel, R. Lewicki, "Tunable mid-infrared laser absorption spectroscopy." Woodhead Publishing Ltd., (2013)

[Tixier-2003a] S. Tixier *et al.* "Surfactant enhanced growth of GaNAs and InGaNAs using bismuth" J. Cryst. Growth, vol. 251 (1-4), pp. 449-454, 2003

[Tixier-2003b] S. Tixier *et al.* "Molecular beam epitaxy growth of GaAs$_{1-x}$Bi$_x$" Appl. Phys. Lett., vol. 82 (14), pp.2245-2247, 2003

[Tournié et Baranov, 2012] E. Tournié and A.N. Baranov, Mid-Infrared lasers: a review, in *Advances in Semiconductor Lasers*, edited by J.J. Coleman, A.C. Brice and C. Jagadish, in Semiconductors and Semimetals, vol. **86**, pp. 183 – 226 (Academic Press, 2012).

[Trampert-2004] A. Trampert et al. "Correlation between interface structure and light emission at 1.3–1.55 µm of (Ga, In)(N, As) diluted nitride heterostructures on GaAs substrates" J. Vac. Sci. Technol. B 22, 2195 (2004).

[Volz-2005] K. Volz et al. "Detection of nanometer-sized strain fields in (GaIn)(NAs) alloys by specific dark field transmission electron microscopic imaging" J. Appl. Phys. 97, 014306 (2005).

[Vurgaftman-2005] I. Vurgaftman *et al.* "Band parameters for III-V compound semiconductors and their alloys" J. Appl. Phys. 89, 5815 (2001).

[Vurgaftman-2015] I. Vurgaftman *et al.* "Interband cascade lasers" 2015 J. Phys. D: Appl. Phys. 48 123001

[Wagener-2000] M. C. Wagener *et al.* "Characterization of secondary phases formed during MOVPE growth of InSbBi mixed crystals", J. Cryst. Growth, vol. 213 (1-2), pp. 51-56, 2000

[Wang-2017] L. Wang et al. "Novel Dilute Bismide, Epitaxy, Physical Properties and Device Application" Crystals 2017, 7(3), 63

[Wei-2002] Y. Wei *et al.* "Type II InAs/GaSb superlattice photovoltaic detectors with cutoff wavelength approaching 32 µm" Appl. Phys. Lett. 81, 3675 (2002)

[Willer-2006] U. Willer *et al.* "Near- and mid-infrared laser monitoring of industrial processes, environment and security applications" Opt. Lasers Eng. (2006), 44(7): 699

[Winnewisser-1994] G. Winnewisser, "Submillimeter and infrared astronomy", Infrared Physics and Technology, Vol 35, N°2/3, p 551 (1994)

[Wood-1982] C. E. C. Wood *et al.* "Magnesium- and calcium-doping behavior in molecular-beam epitaxial III-V compounds" J. Appl. Phys.53, 4230 (1982).

[Wood-2016] A.W. Wood et al. "Droplet-mediated formation of embedded GaAs nanowires in MBE GaAs$_{1-x}$Bi$_x$ films" Nanotechnology 27, 115704 (2016).

[Wood-2017] A. W. Wood et al. "Annealing-induced precipitate formation behavior in MOVPE-grown GaAs$_{1-x}$Bi$_x$ explored by atom probe tomography and HAADF-STEM" Nanotechnology 28, 215704 (2017).

[Wu-2014a] M. Wu et al. "Formation and phase transformation of Bi-containing QD-like clusters in annealed GaAsBi" Nanotechnology 25, 205605, (2014).

[Wu-2014b] M. Wu et al. "Observation of atomic ordering of triple-period-A and-B type in GaAsBi" Appl. Phys. Lett. 105 041602, (2014).

[Wu-2015] M. Wu et al. "Detecting lateral composition modulation in dilute Ga(As, Bi) epilayers" Nanotechnology 26, 425701, (2015).

[Wu-2017] X. Wu *et al.*, "1.142 µm GaAsBi/GaAs Quantum Well Lasers Grown by Molecular Beam Epitaxy", ACS Photonics 4, 1322 (2017)

[Yoshimoto-03] M. Yoshimoto *et al.* "Metastable GaAsBi alloy grown by molecular beam epitaxy" Jap. J. of Appl. Phys., vol. 42 (10B), pp. L1235-L1237, 2003

[Yue-2018a] L. Yue *et al.* "Molecular beam epitaxy growth and optical properties of high bismuth content GaSb1-xBix thin films", J. Alloys Compd. 742, 780 (2018)

[Yue-2018b] L. Yue *et al.* "Structural and optical properties of GaSbBi/GaSb quantum wells" Opt. Mat. Express 8, 893-900 (2018).

[Zhang-2018] Y. C. Zhang *et al.* "Wavelength extension in GaSbBi quantum wells using delta-doping" Journal of Alloys and Compounds 744, 667-671 (2018).